\documentclass[11pt, authoryear]{elsarticle}              
\journal{arXiv.org}            

\usepackage{amssymb,amsfonts,amsmath}
\usepackage{xcolor}
\usepackage{natbib}
\usepackage[top=1.1in, bottom=1.0in, left=1.0in, right=1.0 in]{geometry}   
\usepackage[singlespacing]{setspace}
\usepackage[bottom]{footmisc}
\usepackage{indentfirst}
\usepackage{endnotes}
\usepackage{verbatim}                
\usepackage{lastpage,placeins}
\usepackage{hyperref}                
\usepackage{epsfig,pgf,subfig}       
\usepackage{array}
\usepackage{rotating,longtable,booktabs, colortbl}       
\usepackage{mdwlist}
\usepackage{enumerate}
\usepackage[shortlabels]{enumitem}
\usepackage{pdflscape,afterpage}                    
\usepackage{appendix}
\usepackage{bm}
\usepackage{breakcites}

\usepackage[standard]{ntheorem}
\usepackage[algo2e]{algorithm2e}

\usepackage[T1]{fontenc}
\usepackage{lmodern}

\allowdisplaybreaks[1]                     
\theoremstyle{break}
\theorembodyfont{\normalfont}
\newtheorem{algorithm}[algocf]{Algorithm}

\makeatletter    
\def\@author#1{\g@addto@macro\elsauthors{\normalsize%
    \def\baselinestretch{1}%
    \upshape\authorsep#1\unskip\textsuperscript{%
      \ifx\@fnmark\@empty\else\unskip\sep\@fnmark\let\sep=,\fi
      \ifx\@corref\@empty\else\unskip\sep\@corref\let\sep=,\fi
      }%
    \def\authorsep{\unskip,\space}%
    \global\let\@fnmark\@empty
    \global\let\@corref\@empty  
    \global\let\sep\@empty}%
    \@eadauthor={#1}
}
\makeatother     

\begin{document}

\begin{frontmatter}
\title{\textbf{Seemingly Unrelated Regression with Measurement Error: Estimation
via Markov chain Monte Carlo and Mean Field Variational Bayes Approximation}
}

\author{Georges Bresson\corref{cor1}}\ead{georges.bresson@u-paris2.fr}
\address{Department of Economics, Universit\'{e} Paris II, France}

\author{Anoop Chaturvedi\corref{cor2}} \ead{anoopchaturv@gmail.com}
\address{Department of Statistics, University of Allahabad, India.}

\author{Mohammad Arshad Rahman\corref{cor3}}\ead{marshad@iitk.ac.in}
\address{Department of Economic Sciences, Indian Institute of Technology Kanpur, India.}

\author{Shalabh\corref{cor4}}\ead{shalab@iitk.ac.in}
\address{Department of Mathematics and Statistics,
Indian Institute of Technology Kanpur, India.}

\cortext[cor3]{Corresponding author}

\fntext[fn1]{We dedicate this article to the memory of Viren K. Srivastava.
We thank the editor Antoine Chambaz, associate editor Laura Sangalli and an
anonymous referee for their valuable comments. We are also grateful to David
Brownstone, Ivan Jeliazkov, Dale Poirier and the participants of the research
seminar (2015) at the University of California, Irvine for a variety of
helpful comments and suggestions on an earlier version.}

\begin{abstract}
Linear regression with measurement error in the covariates is a heavily
studied topic, however, the statistics/econometrics literature is almost
silent to estimating a multi-equation model with measurement error. This
paper considers a seemingly unrelated regression model with measurement
error in the covariates and introduces two novel estimation methods: a pure
Bayesian algorithm (based on Markov chain Monte Carlo techniques) and its
mean field variational Bayes (MFVB) approximation. The MFVB method has the
added advantage of being computationally fast and can handle big data. An
issue pertinent to measurement error models is parameter identification,
and this is resolved by employing a prior distribution on the measurement
error variance. The methods are shown to perform well in multiple
simulation studies, where we analyze the impact on posterior estimates
arising due to different values of reliability ratio or variance of the
true unobserved quantity used in the data generating process. The paper
further implements the proposed algorithms in an application drawn from the
health literature and shows that modeling measurement error in the data can
improve model fitting.

\end{abstract}

\end{frontmatter}



\section{Introduction}

The seemingly unrelated regression (SUR) model consists of a system of linear
multiple regression equations such that each equation has a different
continuous dependent variable with a potentially different set of exogenous
explanatory variables (covariates) and the errors are correlated across
equations \citep{Zellner-1962}. When the conditions of the SUR model apply,
estimators obtained from SUR are more efficient relative to ordinary least
squares estimators. The optimality feature and other theoretical properties
of the SUR estimator within the frequentist framework are well studied in
\citet{Srivastava-Dwivedi-1979}, \citet{Srivastava-Giles-1987} and
\citet{Fiebig-2001}. The Bayesian approach to estimating SUR model was
introduced in \citet{Zellner-1971}, where the author analytically derived the
conditional posterior densities of the parameters. Given the conditional
posteriors, the model can then be estimated using a Markov chain Monte Carlo
(MCMC) technique, known as Gibbs sampling \citep{Geman-Geman-1984,
Casella-George-1992}. Since the introduction in \citet{Zellner-1971}, the
literature on Bayesian analysis of SUR has grown considerably in various
directions, including estimation \textit{via} MCMC
\citep{Percy-1992,Griffiths-Chotikapanich-1997,Griffiths-Valenzuela-2006} and
direct Monte Carlo approach \citep{Zellner-Ando-2010,Ando-Zellner-2010},
prediction in SUR model \citep{Percy-1992} and several model extensions that
include restricted SUR \citep{Steel-1992}, SUR with serially correlated
errors and time varying parameters \citep{Chib-Greenberg-1995} and
semiparametric inference in SUR model \citep{Koop-etal-2005}.

The existing literature on SUR models including the quoted articles have
worked based on the assumption that the covariates are measured correctly.
Nonetheless, in practice there can emerge situations where one or more of the
covariates are recorded with error, thus giving rise to SUR with measurement
error (hereafter SURME). Modeling measurement error within a SUR structure or
more generally in a multi-equation system has largely gone unnoticed in the
literature (both frequentist and Bayesian), the only exception is
\citet{Carroll-etal-Book-2006,Carroll-etal-2006} explained in the next
paragraph. In contrast, there has been considerable work on single equation
models with measurement error. Within a linear regression framework, it is
well known that measurement error in the data leads to bias and inconsistency
in ordinary least squares (OLS) estimator (see for instance
\citet{Cheng-VanNess-1999}, \citet{Fuller-1987},
\citet{Wansbeek-Meijer-2000}, \citet{Rao-etal-2008} and
\citet{Hu-Wansbeek-2017}). To achieve consistency of OLS estimator, side
assumptions are required such as known
measurement error variance or known \textit{reliability ratio}.\footnote{%
If $w$ and $z$ are two random variables such that $w=z+u$ and the error $u$
is independent of $z$, then the reliability ratio $R_{z}$ is defined as the
true variance divided by the total variance, \textit{i.e.}, $%
R_{z}=Var(z)/(Var(z)+Var(u))$. By definition $0\leq R_{z}\leq 1$.} However,
consistent estimator of regression parameters without the side assumptions
can be constructed when measurement errors have replicated observations
\citep{Shalabh-2003}. Measurement error in nonlinear models is discussed in
\citet{Carroll-etal-2006} along with the Bayesian analysis of linear and
non-linear measurement error models.

Within the multi-equation framework, \citet{Carroll-etal-2006} consider a
combination of linear mixed measurement error model and SUR model to
understand the properties of measurement error in food frequency
questionnaire data for protein and energy. They adopt the frequentist
estimation approach and use a nearby adaptive method based on weighted Akaike
information criterion (AIC) to select the best fitting model, a form of model
averaging which is popular in the Bayesian literature.
\citet{Carroll-etal-2006} find that a fully parameterized model in which
measurement errors in the two nutrients are modeled jointly, offers no gain
in efficiency compared to fitting each model separately. However, when some
parameters are set to zero resulting in a reduced model, considerable gains
in efficiency is attained. We may adopt the frequentist approach to
estimating SURME model with \emph{structural} measurement error, but it is
fraught with difficulty because the number of parameters become larger than
the number of normal equations derived from the likelihood function. In such
cases, side assumptions can be used to identify the model as done in linear
regression, but even then deriving the maximum likelihood estimators for
SURME model is a challenging task. Besides, ignoring measurement error in the
data can lead to a poor model fit.

In this paper, we introduce two novel methods---a pure Bayesian algorithm and
a mean field variational Bayes (MFVB) technique---to estimate the SURME model
where each equation can potentially have a different covariate that is
measured with error. Both the approach employs a classical structural form of
measurement error and the link between the covariate measured with error and
the other covariates (with no measurement error) is modeled through an
exposure equation. Identification of parameters is achieved by placing a
prior distribution on the measurement error variance. The pure Bayesian
approach is analytically simpler and produces tractable conditional
distributions which enables the use of Gibbs sampling. However, the MCMC
draws of the parameters corresponding to the covariate measured with error
tend to be highly correlated. To reduce autocorrelation in MCMC draws, one
may consider \emph{thinning} i.e., use every $l$-th draw in estimating the
parameter. Thinning is debatable and while some authors such as
\citet{Owen-2017} recommend thinning, others such as \citet{Link-Eaton-2012}
advise against the use of thinning. So, we explore other methods and come up
with a more elegant solution to the problem of high autocorrelation i.e., the
MFVB approach to estimating SURME model.

We illustrate both the techniques in multiple simulation studies and compare
the results to a standard SUR model, where we ignore or do not model the
measurement error. In the first set of simulation studies, data are generated
from a SURME model using different values of the variance of the true
unobserved variable, while holding the reliability ratio fixed. In the second
set of simulations, data are generated using different values of reliability
ratio, holding the variance of the true unobserved variable at a fixed value.
The results suggest that both the proposed methods perform well and correctly
highlight the importance of modeling measurement error within the SUR
structure when variables are measured with error. In addition, the SURME
model is implemented in an application drawn from the health literature and
estimated using the two proposed methods. Specifically, weight and high
density lipoprotein (\emph{HDL}) are jointly modeled as a function of several
covariates and blood pressure, which is common to both equations and
considered to have measurement error. Blood pressure is modeled as a function
of the covariates in the exposure equation. Model selection exemplify the
practical utility of the SURME model compared to the standard SUR model.

The remainder of the paper is organized as follows. Section 2 presents the
SURME model, derives the joint posterior density and proposes a Gibbs
sampling algorithm to estimate the model. Section 3 develops the MFVB
approximation of the MCMC algorithm. Section 4 demonstrates the two
algorithms in several Monte Carlo simulation exercises and Section 5 presents
an application drawn from health literature. Section 6 concludes.

\section{The SURME Model and Estimation via Gibbs sampling}

The seemingly unrelated regression with measurement error (SURME) model
incorporates measurement error for covariates in the SUR model and can be
expressed in terms of the following equations,
\begin{equation}
y_{mi}=x_{mi}^{\prime }\beta _{m}+z_{mi}\gamma _{m}+\varepsilon _{mi},
\qquad m=1,...,M; \; i=1,...,N,  \label{sec2:eq_1}
\end{equation}
where the response $y_{mi}$ is a scalar, $x_{mi}^{\prime }$ is $\left(
1\times k_{m}\right) $ vector of covariates, $z_{mi}$ is a true unobserved
scalar covariate that is prone to measurement error, and the subscripts $m$
and $i$ denote the equation number and individual/observation, respectively.
Stacking the equations for each $i$, we can write model (\ref{sec2:eq_1}) as
follows,
\begin{equation}
y_{i}=X_{i}\beta +Z_{i}\gamma +\varepsilon _{i}, \qquad i=1,..,N,
\label{sec2:eq_2}
\end{equation}
where $y_{i}=\left( y_{1i},...,y_{Mi}\right) ^{\prime }$ and $\gamma =\left(
\gamma _{1},...,\gamma _{M}\right) ^{\prime }$ are vectors of dimension
$\left( M\times 1\right)$, and $\beta =\left( \beta _{1},...,\beta
_{M}\right) ^{\prime }$  is of dimension $(K \times 1)$, where $K =
k_{1}+\cdots + k_{M}$. The matrices,
\begin{equation*}
X_{i}=\left(
\begin{array}{ccc}
x_{1i}^{\prime } & \cdots & 0 \\
&   \ddots &  \\
0 & \cdots & x_{Mi}^{\prime }%
\end{array}%
\right)
\qquad \mathrm{and} \qquad
Z_{i}=\left(
\begin{array}{cccc}
z_{1i}  & \cdots & 0 \\
 & \ddots &  \\
0  & \cdots & z_{Mi}
\end{array}%
\right),
\end{equation*}
are of dimension $(M \times K)$ and $(M \times M)$, respectively. In
addition, the error $\varepsilon_{i}$ is assumed to be independently and
identically distributed (\emph{i.i.d.}) as a normal distribution i.e.,
$\varepsilon_{i} \sim N(0, \Sigma_{\varepsilon})$ for $i=1,\cdots,N$, where
the covariance,
\begin{displaymath}
\Sigma_{\varepsilon} = \left( \begin{array}{ccc}
\sigma_{11} &  \cdots   &   \sigma_{1M}     \\
\vdots      &  \ddots   &   \vdots          \\
\sigma_{M1} &  \cdots   &   \sigma_{MM}
\end{array} \right),
\end{displaymath}
is a symmetric matrix that permits nonzero correlation across equations (or
first subscript) for any given individual (or second subscript) and ties each
independent regression into a system of equations, hence the phrase seemingly
unrelated regression. Measurement error in reference to model
(\ref{sec2:eq_2}) arises because $Z_{i}$ is not observed, instead we observe
$W_{i}$ which is a sum of the true unobserved quantity $Z_{i}$ and a
measurement error term $u_{i}$. This definition implies a \emph{classical
measurement error} \citep{Fuller-1987}. Additionally, we assume that the true
unobserved quantity $Z_{i}$ follows a distribution, so that the measurement
error model is of the \emph{structural form}. This can be represented as
follows,
\begin{equation}
\widetilde{W}_{i}=\widetilde{Z}_{i}+\widetilde{u}_{i}, \qquad
\widetilde{u}_{i}\sim N_{M}\left( 0,\sigma _{u}^{2}I_{M}\right), \qquad
\textit{classical structural form}, \label{sec2:eq_3}
\end{equation}
where for algebraic simplification, we use the notations
$\widetilde{W}_{i}=\left( w_{1i},...,w_{Mi}\right) ^{\prime }$,
$\widetilde{Z}_{i}=\left( z_{1i},...,z_{Mi}\right) ^{\prime }$,
$\widetilde{u}_{i}=\left( u_{1i},...,u_{Mi}\right) ^{\prime }$, then
$W_{i}=diag(\widetilde{W}_{i})$, $ Z_{i}=diag(\widetilde{Z}_{i})$,
$u_{i}=diag(\widetilde{u}_{i})$ are $\left( M\times M\right) $ diagonal
matrices and $I_{M}$ is a ($M \times M$) identity matrix.

An interesting addition to equation~\eqref{sec2:eq_3} is to relate the
primary explanatory variable of interest (here $Z_{i}$) to other covariates
($X_{i}$), giving rise to the \emph{exposure model}. The term ``exposure
model'' comes from epidemiology, where the primary explanatory variable is
affected by exposure to ``toxicants'' or ``risk factors''. Therefore, the
potential links between the latent variable $Z$ and the other covariates $X$
can be expressed as follows,
\begin{equation}
\widetilde{Z}_{i}= X_{i} \omega + \widetilde{\varepsilon}_{z,i}, \qquad
\widetilde{\varepsilon}_{z,i} \sim N_{M}\left( 0, \sigma _{Z}^{2}I_{M} \right),
\qquad  \textit{exposure model}. \label{sec2:eq_4}
\end{equation}
The three equations~\eqref{sec2:eq_2}, \eqref{sec2:eq_3} and
\eqref{sec2:eq_4} together define our SURME model and the resulting
likelihood is derived as follows,
\begin{align}
\begin{split}
& f(y,W,Z|X,\Delta) = \prod_{i=1}^{N} f(y_{i}, W_{i}, Z_{i}|X, \Delta )\\
& = \prod_{i=1}^{N} \bigg\{ f(y_{i}| W_{i}, Z_{i},X, \Delta ) \times
    f(W_{i}|Z_{i}, X,\Delta) \times f(Z_{i}|X,\Delta) \bigg\}\\
& = \prod_{i=1}^{N} \bigg\{ f(y_{i}| Z_{i},X, \Delta ) \times
    f(W_{i}|Z_{i}, X,\Delta) \times f(Z_{i}|X,\Delta) \bigg\}\\
& = \prod_{i=1}^{N} \bigg\{ (2\pi)^{-M/2} \, |\Sigma_{\varepsilon}|^{-1/2}
    \exp\Big[-\frac{1}{2} (y_{i} - X_{i}\beta - Z_{i}\gamma)'
    \Sigma_{\varepsilon}^{-1}(y_{i} - X_{i}\beta - Z_{i}\gamma)   \Big] \\
& \hspace{0.53in} \times (2\pi)^{-M/2} \, (\sigma_{u}^{2})^{-M/2}
    \exp\Big[-\frac{1}{2 \sigma_{u}^{2}} (\widetilde{W}_{i} - \widetilde{Z}_{i})'
    (\widetilde{W}_{i} - \widetilde{Z}_{i}) \Big] \\
& \hspace{0.53in} \times (2\pi)^{-M/2} \, (\sigma_{Z}^{2})^{-M/2}
    \exp\Big[-\frac{1}{2 \sigma_{Z}^{2}} (\widetilde{Z}_{i} - X_{i}\omega)'
    (\widetilde{Z}_{i} - X_{i}\omega) \Big]  \bigg\},
\end{split}
\label{sec2:eq_5}
\end{align}
where $\Delta \equiv (\beta, \gamma, \Sigma_{\varepsilon}, \omega,
\sigma_{Z}^{2},\sigma_{u}^{2})$ and as mentioned earlier, $\widetilde{W}_{i}$
and $\widetilde{Z}_{i}$ are column vectors that contain the diagonal elements
of the matrices $W_{i}$ and $Z_{i}$, respectively.

Before proceeding with estimation, we add a few words on identification
issues that typically arise with measurement error models. In linear
regression with measurement error, identification of parameters require
additional assumptions. Such assumptions can be constant measurement error
variance, known reliability ratio or some other conditions as presented in
\citet{Cheng-VanNess-1999}. The same identification conditions are also
applicable to the proposed SURME model under the existing distributional
assumptions. Nonetheless, we follow a purely Bayesian approach and employ
prior distributions to identify the parameters of the model \citep[see][Chap.
5]{Zellner-1971}.

The Bayesian estimation method combines the likelihood of the model with
suitable prior distributions to obtain the joint posterior distribution. We
utilize the following prior distributions:
\begin{equation}
\begin{split}
& \beta  \sim  N_{K} \left( \beta_{0},B_{0}\right), \quad
\gamma \sim  N_{M}\left( \gamma_{0}, G_{0}\right), \quad
\Sigma_{\varepsilon}^{-1} \sim W_{M} \left( \nu_{0}, S_{0}\right), \\
& \omega \sim N_{K}\left( \omega _{0},O_{0}\right), \quad
\sigma_{Z}^{2}  \sim IG\left( \delta _{1},\delta _{2}\right), \quad
\sigma_{u}^{2}  \sim IG\left( \delta _{3},\delta _{4}\right),
\end{split}
\label{sec2:eq_6}
\end{equation}
where $W_{M}$ denotes a Wishart distribution of dimension $M$ and $IG$
denotes an inverse gamma distribution. Here we note that if one is not
interested in the exposure equation, it can be dropped from the model. In
such a case, $\tilde{Z}_{i} \sim N(\mu, \sigma_{Z}^{2} I_{M})$ and $\mu$ can
be given a normal prior as $\mu \sim N(\mu_{0}, \sigma_{\mu}^{2} I_{M})$.
Coming back to the SURME model, the joint posterior distribution can be
obtained by combining the likelihood \eqref{sec2:eq_5} with the prior
distributions \eqref{sec2:eq_6} as follows,
\begin{allowdisplaybreaks}
\begin{equation}
\begin{split}
p\left(\Delta,Z \vert y, X, W\right) & \propto
\prod_{i=1}^{N} \Bigg\{ \left\vert
\Sigma_{\varepsilon}\right\vert^{-1/2} \exp \left[
-\frac{1}{2}\left( y_{i}-X_{i}\beta -Z_{i}\gamma \right)^{\prime}
\Sigma_{\varepsilon}^{-1}\left(y_{i} - X_{i}\beta - Z_{i}\gamma
\right) \right]  \\
& \quad \times \left( \sigma_{u}^{2}\right)^{-M/2}\exp
\left[ -\frac{1}{2\sigma_{u}^{2}}\left( \widetilde{W}_{i} -
\widetilde{Z}_{i}\right)^{\prime}\left( \widetilde{W}_{i}-\widetilde{Z}_{i}\right) %
\right] \\
& \quad \times \left( \sigma_{Z}^{2}\right)^{-M/2}\exp
\left[ -\frac{1}{2\sigma_{Z}^{2}}\left( \widetilde{Z}_{i}- X_{i}
\omega \right)^{\prime} \left( \widetilde{Z}_{i} - X_{i} \omega \right)
\right] \Bigg\} \\
& \quad \times \left\vert
B_0\right\vert^{-1/2} \exp \left[
-\frac{1}{2}\left( \beta -\beta_{0} \right)^{\prime}
B_{0}^{-1}\left(\beta - \beta_{0}\right) \right]  \\
& \quad \times \left\vert
G_0\right\vert^{-1/2} \exp \left[
-\frac{1}{2}\left( \gamma -\gamma_{0} \right)^{\prime}
G_{0}^{-1}\left(\gamma - \gamma_{0}\right) \right]  \\
& \quad \times \left\vert
\Sigma_{\varepsilon}\right\vert^{-\frac{s_0 -M-1}{2}} \exp \left[
-\frac{1}{2} \text{tr} \left( S_{0}^{-1} \Sigma^{-1}_{\varepsilon} \right) \right]  \\
& \quad \times \left( \sigma_{Z}^{2}\right)^{-\delta_1 - 1}\exp
\left[ -\frac{\delta_2}{ \sigma_{Z}^{2}} \right]  \times \left( \sigma_{u}^{2}\right)^{-\delta_3 - 1}\exp
\left[ -\frac{\delta_4}{ \sigma_{u}^{2}} \right].
\end{split}
\label{sec2:eq_7}
\end{equation}
\end{allowdisplaybreaks}
Typical with the Bayesian approach, the joint posterior density
(\ref{sec2:eq_7}) is not tractable and the parameters are sampled using MCMC
techniques. To this purpose, conditional posterior densities of the
parameters are derived (see Appendix A in the supplementary material) and
Gibbs sampling is employed to estimate the model as exhibited in
Algorithm~\ref{alg:algorithm1}. Note that some of the conditional posteriors
are conditioned on a subset of parameters, but these are full conditionals
that just do not depend on the full set of parameters. Conditional posteriors
that depend on a subset of parameters have also been referred to as reduced
conditional posteriors and Gibbs sampling as partially collapsed Gibbs
sampling \citep[see][]{Liu-1994,vanDyk-Park-2008}.

\begin{table*}[!t]
\begin{algorithm}[Gibbs sampling for SURME model]
\label{alg:algorithm1} \rule{\textwidth}{0.5pt}
\begin{enumerate}
\item Sample $\beta|\gamma, \Sigma_{\varepsilon},Z,y \sim N_{K}\left( \overline{\beta },B_{1}\right)$,
    where,
    \newline
$B_{1}^{-1}=\left[ \displaystyle \sum_{i=1}^{N} X'_{i} \Sigma _{\varepsilon
}^{-1}X_{i}+B_{0}^{-1}\right] $, $\overline{\beta }=B_{1}\left[
\displaystyle \sum_{i=1}^{N} X_{i}^{\prime }\Sigma _{\varepsilon
}^{-1}y_{i}^{\ast }+B_{0}^{-1}\beta _{0}\right]$,  and $y_{i}^{\ast
}=y_{i}-Z_{i}\gamma $.
\item Sample $\gamma|\beta,\Sigma_{\varepsilon},Z,y \sim
    N_{M}\left( \overline{\gamma },G_{1}\right) $,
    where, \newline
$G_{1}^{-1}=\left[ \displaystyle \sum_{i=1}^{N} Z_{i}^{\prime}\Sigma
_{\varepsilon }^{-1}Z_{i}+G_{0}^{-1} \right] $,
$\overline{\gamma}=G_{1}\left[ \displaystyle \sum_{i=1}^{N} Z_{i}^{\prime}
\Sigma _{\varepsilon }^{-1}\tilde{y}_{i} + G_{0}^{-1} \gamma _{0} \right]$,
and $\tilde{y}_{i}=y_{i}-X_{i}\beta$.
\item Sample $\Sigma_{\varepsilon}^{-1}| \beta, \gamma, Z,y \sim W_{M}
    \left(\nu_{1},S_{1}\right) $, where, \newline $\nu_{1}=\nu_{0}+N$ and
    $S_{1}^{-1}=\left[ S_{0}^{-1} + \displaystyle \sum_{i=1}^{N}\left(
    y_{i}-X_{i}\beta -Z_{i}\gamma \right) \left( y_{i}-X_{i}\beta
    -Z_{i}\gamma \right)^{\prime }\right]$.
\item Sample $\tilde{Z}_{i}|\beta,\gamma,\Sigma _{\varepsilon},\omega,
    \sigma_{Z}^{2},\sigma_{u}^{2},W,y
    \sim N_{M}\left( M_{1,i},M_{2}\right)$, $\forall
    i=1,...,N$, where, \newline $M_{2}^{-1}=\left[ \Psi +\left(
    \frac{1}{\sigma_{Z}^{2}}+\frac{1}{\sigma_{u}^{2}}\right)
    I_{M}\right]$, with $\Psi =\Gamma \odot \Sigma_{\varepsilon }^{-1}$,
    $\Gamma =\gamma \gamma^{\prime }$, \newline and $M_{1,i}=M_{2}\left[
    diag(\gamma )\Sigma_{\varepsilon}^{-1}\left(y_{i}-X_{i}\beta \right) +
    \frac{\widetilde{W}_{i}}{\sigma_{u}^{2}} + \frac{X_{i} \omega}
    {\sigma_{Z}^{2}} \right]$, where $\odot $ is the dot (or
    Hadamard) product.
\item Sample $\omega|Z,\sigma_{Z}^{2} \sim
    N_{M}\left( \omega_1, {\Sigma_{\omega} }\right)
    $, where, \newline $\Sigma_{\omega}^{-1} = \left[  \frac{1}{\sigma
    _{Z}^{2}} \displaystyle \sum_{i=1}^{N} X^{\prime}_{i} X_{i}  + O^{-1}_0
    \right] $ and $\omega_1= \Sigma_{\omega} \left[ \frac{1}{\sigma
    _{Z}^{2}} \displaystyle \sum_{i=1}^{N} X^{\prime}_i \tilde{Z}_{i} +
    O^{-1}_0 \omega_0 \right] $.
\item Sample $\sigma_{Z}^{2}|Z,\omega \sim IG\left( \delta_{1}^{\ast },
    \delta_{2}^{*}\right) $, where, \newline $\delta_{1}^{\ast} = \delta_{1}
    +\frac{N.M}{2}$ and $\delta_{2}^{\ast}=\delta_{2}+\frac{1}{2}
    \displaystyle \sum_{i=1}^{N} \left( \tilde{Z}_{i}- X_{i} \omega
    \right)^{\prime}\left( \tilde{Z}_{i}- X_{i} \omega \right) $.
\item Sample $\sigma _{u}^{2}|Z,W \sim IG\left(\delta_{3}^{\ast},
    \delta_{4}^{\ast}\right)$, where, \newline $\delta_{3}^{\ast}
    = \delta_{3}+\frac{N.M}{2}$ and $\delta_{4}^{\ast}
    =\delta_{4}+\frac{1}{2}\displaystyle \sum_{i=1}^{N}
    \left(\tilde{W}_{i}-\tilde{Z}_{i}\right)^{\prime} \left(
    \tilde{W}_{i}-\tilde{Z}_{i}\right) $.
\end{enumerate}
\rule{\textwidth}{0.5pt}
\end{algorithm}
\end{table*}

The sampling algorithm, presented in Algorithm~\ref{alg:algorithm1}, shows
that $\beta$ and $\gamma$ are sampled from an updated multivariate normal
distribution. Standard result is obtained for the precision matrix
$\Sigma_{\varepsilon}^{-1}$, which is sampled from an updated Wishart
distribution. All the three parameters ($\beta,\gamma,\Sigma_{\varepsilon}$)
follow their respective distributions marginally of $\omega$,
$\sigma_{u}^{2}$ and $\sigma _{Z}^{2}$. The true unobserved quantity $Z$ is
drawn from an updated multivariate normal distribution conditional on all the
remaining model parameters. Similarly, $\omega $ is sampled from an updated
multivariate normal distribution conditional on $\left( Z,\sigma
_{Z}^{2}\right) $. The two variance parameters are drawn from updated inverse
gamma distributions with $\sigma _{Z}^{2}$ conditioned on $\left(Z,\omega
\right) $ and $\sigma _{u}^{2}$ conditioned on $\left(W, Z \right)$. Note
that if we drop the exposure equation from the SURME model,
Algorithm~\ref{alg:algorithm1} only requires a slight modification. In this
context, $\mu$ replaces $X_{i}\omega$ and is sampled from an updated normal
distribution as $\mu|Z,\sigma_{Z}^{2} \sim N_{M}(\bar{l}, \bar{\Lambda})$,
where $\bar{l} = \bar{\Lambda} \Big( \sum_{i=1}^{N} \widetilde{Z}_{i} /
\sigma_{Z}^{2} + \frac{\mu_{0}}{\sigma_{\mu}^{2}} \Big)$ and
$\bar{\Lambda}^{-1} =  \Big( \frac{N}{\sigma_{Z}^{2}} +
\frac{1}{\sigma_{\mu}^{2}}\Big) I_{M}$ are the posterior mean and posterior
precision, respectively.

We note that the model presented in this paper utilizes the structural
measurement error model which assumes that $Z$ follows a distribution. Hence,
the distribution of $Z$ was introduced as a part of the model. However, in
the measurement error literature, there is another form of measurement error
known as \emph{functional} form. The functional measurement error model
assumes that the true unobserved quantity $Z$ is fixed. In our modeling and
estimation framework, we can easily incorporate the functional form of
measurement error by modeling the distribution of $Z$ as a part of the
subjective prior information \citep{Zellner-1971}. This implies that the
joint posterior distribution \eqref{sec2:eq_6} will be unchanged and
derivations of the conditional posterior distributions will proceed in
exactly the same way as described in Appendix~A of the supplementary file. To
reiterate, the fundamental difference in analyzing SUR model with structural
and functional forms of measurement error lies in the interpretation given to
the distribution of $Z$, the true unobserved quantity.

In the MCMC estimation of SURME model, one consideration that arise is that
$Z$ and $\gamma $ are both unknown, and drawing them conditional on each
other lead to high autocorrelation in MCMC draws. This occurrence is a
general problem and happens when two or more unknown variables/parameters
that appear in product form are drawn conditional on each other. To reduce
the autocorrelation in MCMC draws (and consequently reduce the inefficiency
factors) some authors\footnote{See for instance \citet{Jeliazkov-2013} for
the case of latent variables in a non parametric VAR specification.} propose
to improve mixing by sampling $\gamma $ from the marginal distribution and
then sampling $Z|\gamma $ or \textit{vice versa}. However, deriving the
marginal posterior distribution of $\gamma$ (or of $Z$) is not
straightforward and the marginalization trick do not improve the results in
our modeling context. \footnote{Many thanks to Ivan Jeliazkov and the
participants of the UCI seminar for the suggestion to sample $\gamma$
marginally of $Z$ and then sampling $Z|\gamma$. See appendix E in the
supplementary material. However, the several tests we conducted did not
improve our initial results with standard Gibbs sampling.}
As a solution to reduce autocorrelation, many researchers have employed
\emph{thinning} to improve the mixing of the draws. The thinning of MCMC
draws has been criticized by some authors including
\citet{MacEachern-Berliner-1994}, \citet{Link-Eaton-2012}), but others such
as \citet{Geyer-1991} acknowledges that thinning can increase statistical
efficiency. In a recent paper, \citet{Owen-2017} shows that the usual advice
against thinning can be misleading. We employ thinning to improve the mixing
properties of the MCMC draws of $\gamma$ in our simulation studies and
application. However, given the controversy around thinning, we explore other
methods and come up with the MFVB approximation to estimate SURME model.

\section{The mean field variational Bayes (MFVB) approximation}

Variational Bayes is an alternative to MCMC methods that provides a
locally-optimal, exact analytical solution to an approximation of the
posterior distribution. The parameters of the approximate distribution are
selected to minimize the Kullback-Leibler divergence (a distance measure)
between the approximation and the posterior. The MFVB approximation is a
deterministic optimization approach and so is particularly useful for big
data sets and/or models with large sparse covariance matrices. Besides, it is
similar to Gibbs sampling for conjugate models. Some recent articles on MFVB
approach include \citet{Bishop-2006}, \citet{Ormerod-Wand-2010},
\citet{Pham-etal-2013}, \citet{Lee-Wand-2016} and \citet{Blei-etal-2017}.

Suppose, $y$ denotes an observed data vector and $\theta$ is a parameter
vector defined over the parameter space $\Theta$. Following the Bayes
theorem, the posterior distribution can be written as:
\begin{equation*}
p\left(\theta|y\right) =\frac{p\left( \theta,y\right) }{p\left(
y\right) } = \frac{p\left( y|\theta \right) p\left( \theta \right) }{
p\left(y\right) },
\label{sec3:eq_1}
\end{equation*}
where $p\left( y\right) =\int_{\Theta }p\left( \theta ,y\right) d\theta$ is
the marginal likelihood. Let $q$ be an arbitrary density function over
$\Theta $. Then, the logarithm of the marginal likelihood is,
\begin{equation}
\begin{split}
\log p\left(y\right) & = \log p(y) \int_{\Theta} q(\theta) d\theta
= \int_{\Theta} q(\theta) \log p(y) d\theta \\
& = \int_{\Theta} q(\theta) \log \left\{ \frac{p(\theta,y)/q(\theta)}
{p(\theta/y)/q(\theta)}  \right\} d\theta \\
& = \int_{\Theta }q\left( \theta \right) \log \left\{ \frac{p\left( \theta
,y\right) }{q\left( \theta \right) }\right\} d\theta +\int_{\Theta }q\left(
\theta \right) \log \left\{ \frac{q\left( \theta \right) }
{p\left( \theta|y\right) }\right\} d\theta  \\
& = \log \underline{p}\left(y,q\right) +  KL(q,p) \\
&= \log \underline{p}\left(y,q\right) + E_{q\left( \theta \right) }\left[ \log q\left( \theta \right) \right]
-E_{q\left( \theta \right) }\left[ \log p\left( \theta ,y\right) \right]
+ \log p\left( y\right),
\label{sec3:eq_2}
\end{split}
\end{equation}
where $\log \underline{p}\left(y,q\right) = E_{q(\theta)} \left[ \log \left(
\frac{ p\left( \theta, y\right)}{ q\left( \theta \right)} \right) \right]$
denotes the lower bound on the marginal log-likelihood and $KL(q,p)=
E_{q\left( \theta \right) }\left[ \log q\left( \theta \right) \right]
-E_{q\left( \theta \right) }\left[ \log p\left( \theta |y\right) \right]$ is
the Kullback-Leibler divergence  $q\left(\theta \right) $ and $p\left(
\theta|y\right)$. Since $\log p(y)$ is a constant, the minimization of
$KL(q,p)$ is equivalent to maximizing the scalar quantity $\log
\underline{p}\left(y,q\right)$, typically known as evidence lower bound
(ELBO) or variational lower bound. In practice, the maximization of the ELBO
is often preferred to minimization of the KL divergence since it does not
require knowledge of the posterior.

The MFVB approximates the posterior distribution $p\left( \theta|y\right) $
by the product of the $q$-densities,
\begin{equation}
q\left( \theta \right) =\prod_{j=1}^{P}q_{j}\left( \theta _{j}\right).
\label{eq.10}
\end{equation}
Each \textit{optimal} $q$-density  minimizes the Kullback-Leibler divergence
and is given by,
\begin{equation}
q_{j}\left( \theta _{j}\right) \propto \exp \left[ E_{q\left(
-\theta _{j}\right) }\left\{ \log p\left( \theta_{j}| \Omega \right)
\right\} \right] \text{ , }j=1,...,P  \label{eq.11a}
\end{equation}
where $E_{q\left( -\theta _{j}\right)}$ denotes the expectation with respect
to $\prod_{k\neq j}q_{k}\left( \theta _{k}\right)$,  $ \Omega\equiv \left\{
y,\theta _{1},...,\theta _{j-1},\theta _{j+1},...,\theta _{P}\right\} $ is
the set containing all random vectors in the model except $\theta _{j}$, and
$p\left(\theta _{j}|\Omega\right) $ are the full conditional distributions of
the parameters.

For the SURME model, we now consider a MFVB approximation based on the
following factorization:
\begin{equation*}
q\left( \beta ,\gamma ,\Sigma _{\varepsilon },\omega ,\sigma _{Z}^{2},\sigma
_{u}^{2},\tilde{Z}\right) =q\left( \beta \right) q\left( \gamma \right)
q\left( \Sigma _{\varepsilon }\right) q\left( \omega \right) q\left( \sigma
_{Z}^{2}\right) q\left( \sigma _{u}^{2}\right) \prod_{i=1}^{N}q\left( \tilde{%
Z}_{i}\right).  \label{eq.11}
\end{equation*}
These optimal $q$-densities can be derived, as presented in Appendix~B of the
supplementary file, to have the following form,
\begin{equation}
\begin{alignedat}{3}
q\left( \beta \right) & = f_{N_{K}}\left( \mu _{q\left( \beta \right)
},\Sigma _{q\left( \beta \right) }\right)
& \qquad
q\left( \gamma \right) & =  f_{N_{M}}\left( \mu _{q\left( \gamma \right)
},\Sigma _{q\left( \gamma \right) }\right) \\
q\left( \Sigma _{\varepsilon }^{-1}\right) & =  f_{W_{M}}\left( \nu
_{1},B_{q(\Sigma )}\right)
& \qquad
q\left( \omega \right) & =  f_{N_{K}}\left( \mu _{q\left( \omega \right) },\Sigma
_{q\left( \omega \right) }\right) \\
q\left( \tilde{Z}_{i}\right) & =  f_{N_{M}} \left( \mu _{q\left( \tilde{Z}_{i}
\right) },\Sigma _{q\left( \tilde{Z}_{i}\right) }\right)
& \qquad
q\left( \sigma _{Z}^{2}\right) & =  f_{IG}\left( \delta _{1}^{\ast
},B_{q(\sigma _{Z}^{2})}\right) \label{eq.12} \\
q\left( \sigma _{u}^{2}\right) & =  f_{IG} \left( \delta _{3}^{\ast
},B_{q(\sigma _{u}^{2})}\right), &
\end{alignedat}
\end{equation}
where $f$ denotes the density function of the distribution given in the
subscript. The parameters of the optimal densities are updated according to
Algorithm~\ref{alg:algorithm2}. When exposure is dropped, $q(\omega)$ is
replaced with $q(\mu)$ and the optimal density is an updated normal
distribution. Convergence of Algorithm~\ref{alg:algorithm2} is assessed using
the evidence lower bound $\ell$ on the marginal log-likelihood (see
Appendix~C in the supplementary material) that is guaranteed to reach a local
optima based on the convexity property. This algorithm belongs to the family
of coordinate ascent variational inference (CAVI) and iteratively optimizes
each factor of the mean field variational density, while holding the
remaining fixed \citep[see][]{Bishop-2006, Blei-etal-2017}.

\begin{table*}[!t]
\begin{algorithm}[MFVB algorithm for SURME model]
\label{alg:algorithm2} \rule{\textwidth}{0.5pt}
\begin{small}
\begin{enumerate}
\item Initialize $\delta_{1}^{\ast}$, $\delta_{3}^{\ast}$, $B_{q(\sigma
    _{Z}^{2})}$, $B_{q(\sigma_{u}^{2})}$, $\mu_{q\left( \beta \right)}$,
    $\mu_{q\left( \gamma \right)}$, $\mu_{q\left( \omega \right) }$,
    $\Sigma_{q\left( \beta \right)}$, $B_{q(\Sigma)}$, $\Sigma_{q\left(
    \gamma\right)}$, $\Sigma_{q\left( \omega \right)}$, $\mu_{q\left(
    \tilde{Z}_{i}\right) }$, $\Sigma_{q\left( \tilde{Z}_{i}\right) }$ (for
    $i=1,...,N$).
\item Cycle:
\begin{enumerate}
\item $\Sigma_{q\left( \beta \right)} \leftarrow \left[ \sum_{i=1}^{N}
    X_{i}^{\prime }\left( \nu_{1}B_{q(\Sigma )}\right) X_{i} +
    B_{0}^{-1}\right]^{-1}$
\item $\mu_{q\left( \beta \right)} \leftarrow \Sigma_{q\left( \beta \right)}\left[
    \sum_{i=1}^{N} X_{i}^{\prime} \left(\nu_{1}B_{q(\Sigma )}\right)
    \left( y_{i} - diag(\mu_{q\left( \tilde{Z}_{i}\right) }) \mu_{q\left(
    \gamma \right) }\right) + B_{0}^{-1}\beta_{0} \right] $
\item $\Sigma_{q\left( \gamma \right)} \leftarrow \left[ \sum _{i=1}^{N}
    \left( \Sigma_{q\left( \tilde{Z}_{i}\right)} + \mu_{q\left(
    \tilde{Z}_{i}\right)} \mu_{q\left( \tilde{Z}_{i}\right)}^{\prime }
    \right) \odot \left( \nu_{1}B_{q(\Sigma )}\right) + G_{0}^{-1}\right]
    ^{-1}$
\item $\mu_{q\left(\gamma \right)} \leftarrow  \Sigma_{q\left(\gamma \right) }
    \left[\sum_{i=1}^{N} diag(\mu_{q\left( \tilde{Z}_{i}\right)}) \left(
    \nu_{1}B_{q(\Sigma)}\right) \left(y_{i} - X_{i}\mu_{q\left( \beta
    \right) }\right) + G_{0}^{-1}\gamma_{0}\right]$
\item $ B_{q(\Sigma)}  \leftarrow\Bigg[ S_{0}^{-1} + \sum_{i=1}^{N} \bigg[ \left(
    y_{i}-X_{i} \mu_{q\left( \beta \right)} - diag(\mu_{q\left(
\tilde{Z}_{i}\right) }) \mu_{q\left( \gamma \right) }\right) \\
\qquad \times \left( y_{i}-X_{i}\mu_{q\left( \beta \right)} -
diag(\mu_{q\left(\tilde{Z}_{i}\right)}) \mu_{q\left( \gamma \right)
}\right)^{\prime } + X_{i}\Sigma_{q\left( \beta \right) }X_{i}^{\prime}  \\
 \qquad + \left( \mu_{q\left(\tilde{Z}_{i}\right) } \mu_{q\left(
\tilde{Z}_{i}\right)}^{\prime }\right) \odot \Sigma _{q\left( \gamma
\right) } + \Sigma_{q\left( \tilde{Z}_{i}\right)} \odot \left(
\Sigma_{q\left( \gamma \right)} + \mu_{q\left( \gamma \right) }
\mu_{q\left( \gamma \right)}^{\prime }\right) \bigg] \Bigg]^{-1}$
\item $B_{q(\sigma_{Z}^{2})} \leftarrow  \delta_{2} + \frac{1}{2} \sum_{i=1}^{N}
    \left\{ \parallel \mu_{q\left( \tilde{Z}_{i}\right)} - X_{i}
    \mu_{q\left( \omega \right)} \parallel^{2} + \text{tr}\left[
    \Sigma_{q\left( \tilde{Z}_{i}\right)} \right] \right\}$
\item $B_{q(\sigma_{u}^{2})} \leftarrow  \delta_{4} + \frac{1}{2} \sum_{i=1}^{N}
    \left\{ \parallel \tilde{W}_{i} - \mu_{q\left( \tilde{Z}_{i}\right)}
    \parallel^{2} + \text{tr}\left[ \Sigma_{q\left( \tilde{Z}_{i}\right)}
    \right] \right\} $
\item $\Sigma_{q\left( \omega \right) } \leftarrow \bigg[ \left(
    \frac{\delta_{1}^{\ast}}{B_{q(\sigma_{Z}^{2})}}\right)
    \sum_{i=1}^{N}X_{i}^{\prime }X_{i} + O_{0}^{-1}\bigg]^{-1}$
\item $\mu_{q\left( \omega \right)} \leftarrow  \Sigma_{q\left( \omega \right)
    }\left[\left( \frac{\delta_{1}^{\ast}}{B_{q(\sigma_{Z}^{2})}}\right)
    \sum_{i=1}^{N} X_{i}^{\prime }\mu_{q\left( \tilde{Z}_{i}\right)} +
    O_{0}^{-1}\omega_{0}\right] $
\item $\Sigma_{q\left( \tilde{Z}_{i}\right)}  \leftarrow \left[ \left\{
    \Sigma_{q\left( \gamma \right)} + \mu_{q\left(\gamma \right)}
    \mu_{q\left( \gamma \right)}^{\prime}\right\} \odot \left(
    \nu_{1}B_{q(\Sigma)} \right) + \left(
    \frac{\delta_{1}^{\ast}}{B_{q(\sigma_{Z}^{2})}} +
    \frac{\delta_{3}^{\ast}}{B_{q(\sigma_{u}^{2})}}\right)
    I_{M}\right]^{-1}$
\item $\mu_{q\left( \tilde{Z}_{i}\right)}  \leftarrow
    \Sigma_{q\left(\tilde{Z}_{i}\right)} \bigg[ diag(\mu_{q\left( \gamma
    \right) })\left( \nu_{1}B_{q(\Sigma )}\right) \left( y_{i}-X_{i}\mu
    _{q\left( \beta \right)} \right) + \left(
    \frac{\delta_{3}^{\ast}}{B_{q(\sigma_{u}^{2})}}\right) \widetilde{W}_{i} \\
    \quad + \left( \frac{\delta_{1}^{\ast}}{B_{q(\sigma_{Z}^{2})}}
    \right) X_{i}\mu_{q\left( \omega \right) }\bigg]$
\end{enumerate}
\item[]until the increase in the ELBO $(\ell )$ is negligible ($\approx 10^{-7}$).
\end{enumerate}
\end{small}
\rule{\textwidth}{0.5pt}
\end{algorithm}
\end{table*}

The MFVB technique provides computational advantages compared to MCMC because
it is deterministic and does not require a large number of iterations
\citep{Pham-etal-2013,Lee-Wand-2016}. Besides, existing works including
\citet{Bishop-2006}, \citet{Ormerod-Wand-2010}, \citet{Faes-etal-2011},
\citet{Pham-etal-2013}, and \citet{Lee-Wand-2016} suggest that the accuracy
scores of the MFVB approximation, relative to MCMC, generally exceed
$95-97\%$ and rarely goes below $90\%$. Given these advantages, the MFVB
approach can be gainfully utilized for large data models. However, some
authors have reported that covariance matrices from variational approximation
may be typically ``too small'' relative to the sampling distribution of the
maximum likelihood estimator. In this regard, \citet{Blei-etal-2017} opine
that underestimation of the variance should be judged in relation to the task
at hand. However, evidence from empirical research indicate that variational
inference typically do not suffer in accuracy.

\section{Monte Carlo simulation studies}

This section examines the performance of the two proposed methods  in
multiple simulation studies. The first set of simulations (Case I) employ
different values of $\sigma _{Z}^{2}$ to generate the simulated data. The
second set of simulations (Case II) use different values of reliability ratio
defined as $R_{Z}=\sigma _{Z}^{2}/(\sigma _{Z}^{2}+\sigma _{u}^{2})$. In both
sets of simulations, we use a two equation structure represented as follows,
\begin{equation}
\begin{split}
y_{1i} & = \beta_{11}+x_{1i2}\beta_{12} + x_{1i3}\beta_{13} + z_{1i}
\gamma_{1} + \varepsilon_{1i}  \label{eq.m1},  \\
y_{2i} & = \beta_{21}+x_{2i2}\beta_{22} + x_{2i3}\beta_{23} + z_{2i}
\gamma_{2} + \varepsilon_{2i},
\end{split}
\end{equation}
where the first, second and third subscripts in $x_{mij}$ denote the equation
number ($m=1,2$), observation ($i=1,\cdots,N$) and variable number ($j=2,3$),
respectively. The first variable is common to both the equations (i.e.,
$x_{1i2}=x_{2i2}$ for all $i=1,...,N$) and the remaining covariates are
exclusive to the respective equations. Moreover, we assume the error prone
covariate $Z_{i}=diag(z_{1i},z_{2i})$ for all $i$ is unobserved, but is
defined by an exposure model as follows,
\begin{equation}
\begin{split}
z_{1i} & =  \omega _{11}+x_{1i2}\omega _{12}+x_{1i3}\omega_{13}+ \varepsilon _{z1i},
            \label{eq.m2}   \\
z_{2i} & =  \omega_{21}+x_{2i2}\omega_{22}+x_{2i3}\omega_{23}+\varepsilon _{z2i}.
\end{split}
\end{equation}
The unobserved $Z_{i}$ is related to the observed $W_{i}=diag(w_{1i},w_{2i})$
by the equations below,
\begin{equation}
\begin{split}
w_{1i} & =  z_{1i} + u_{1i},   \label{eq.m3}  \\
w_{2i} & =  z_{2i} + u_{2i}.
\end{split}
\end{equation}
Note that the estimation of the SURME model solely relies on $W$ and the role
of $Z$ is limited to generating values for $(W,y)$.

To proceed with data generation, we assign specific values to the parameters
$\beta $, $\gamma $, $\omega$, $\Sigma _{\varepsilon }$, $\sigma _{Z}^{2}$,
$\sigma _{u}^{2}$ and generate $N=300$ observations in each simulation study
for all the variables in the model. Let $\beta _{11}=3$, $\beta _{12}=5
$, $\beta _{13}=4$, $\beta _{21}=4$, $\beta _{22}=3.8$, $\beta _{23}=3$, $%
\gamma _{1}=4$, $\gamma _{2}=4$, $\omega_{11}=1.5$, $\omega_{12}=0.75$,
$\omega_{13}=0.3$, $\omega_{21}=1.5$, $\omega_{22}=1.05$, and
$\omega_{23}=0.45$.
For all values of $i$, the error vector $%
\varepsilon _{i}=\left( \varepsilon _{1i},\varepsilon _{2i}\right) ^{\prime }
$ is generated from a bivariate normal distribution $N(0_{M},\Sigma
_{\varepsilon })$, where $\Sigma _{\varepsilon }=[1$ $0.5$; $0.5$ $1]$.
Values for the common covariate ($x_{1i2}=x_{2i2}$) are generated from
$U(0,2)$ and values for the exclusive covariates $ x_{1i3}$ and $x_{2i3}$ are
generated from $U(0,4)$, where $U$ denotes an uniform distribution. Values
for $\widetilde{Z}_{i}=\left( z_{1i},z_{2i}\right) ^{\prime }$ are generated
as $\widetilde{Z}_{i}\sim N$($X_i\omega ,\sigma_{Z}^{2}I_{M}$), and the
$\widetilde{W}_{i}$'s are generated as $\widetilde{W}
_{i}=\widetilde{Z}_{i}+\widetilde{u}_{i}$ where $\widetilde{u}_{i}\sim
N$($0_{M},\sigma _{u}^{2}I_{M} $). The above setting remains the same in the
following subsections, with change occurring only in values of $\sigma
_{u}^{2}$ (through $R_{Z}$) or $ \sigma _{Z}^{2}$.

In Case I, we investigate the performance of the proposed algorithms in two
simulation studies where the reliability ratio $R_{Z}$ is fixed ($R_{Z}=0.8$)
and $\sigma _{Z}^{2}$ is gradually decreased. Specifically, two values are
considered $\sigma_{Z}^{2}=\{1, 0.0625 \}$. The definition of $R_{Z}$ is used
to generate the corresponding values for
$\sigma_{u}^{2}=\sigma_{Z}^{2}(1-R_{Z})/R_{Z}$, which leads to a
noise-to-true variance ratio $(1-R_{Z})/R_{Z}$ of 25\%. In Case II, we again
examine the performance of the proposed algorithms in two simulation studies
by keeping $\sigma_{Z}^{2}$ fixed ($\sigma_{Z}^{2}=0.0625$) and using two
values of reliability ratio $R_{Z}= \{0.8, \; 0.5714 \} $. The chosen values
are similar to those used in \citet{Pham-etal-2013} and leads to
noise-to-true variance ratios of 25\% and 75\%, respectively. We could define
a noise-to-true variance of $100\%$, $150\%$ or more, but in those cases we
will be dealing more with outliers than with measurement errors.

Bayesian procedures require prior distribution on the parameters of the
model. For the SURME model, we stipulate the following priors: $\beta \sim
N_{K}\left( \beta _{0},B_{0}\right) $ with $ \beta _{0}=\iota _{K}$,
$B_{0}=I_{K}$, and $\iota _{M}$ is a $ \left( M\times 1\right) $ vector of
ones; $\gamma \sim N_{M}\left( \gamma_{0},G_{0}\right) $ with
$\gamma_{0}=\iota_{M}$, $G_{0}=I_{M}$; $\omega \sim N_{K}\left( \omega_{0},
O_{0}\right) $ with $\omega_{0}=\iota_{K}$, $O_{0}=I_{K}$;
$\Sigma_{\varepsilon}^{-1} \sim W_{M}\left( \nu_{0}, S_{0}\right)$ with
$\nu_{0}=50$ and $S_{0}= \nu_{0}[1$ $0.5$; $0.5$ $1]$; $\sigma_{Z}^{2}\sim
IG\left( \delta_{1},\delta_{2}\right) $ and $\sigma_{u}^{2}\sim IG\left(
\delta_{3},\delta _{4}\right)$ with
$\delta_{1}=\delta_{2}=\delta_{3}=\delta_{4}=1/100$. All these priors are
proper yet specify vague information about the parameters mainly for the
measurement error $u_{i}$ and the error prone covariate $Z_{i}$. In addition,
the same priors are used in all the simulations to highlight the effect of
changing $R_{Z}$ or $\sigma _{Z}^{2}$ in estimation of the parameters and
consequently on the performance of the algorithms.

The MCMC results are obtained from $50,000$ draws, after a burn-in of $1,000$
draws. We replicate these simulations $100$ times and report the means over
these $100$ replications of the posterior means\footnote{ To save time, we
only run $51,000$ draws per replication. Higher number of MCMC draws, such as
$100,000$ or $200,000$, exponentially increase the computing time without any
increase in precision. As an example for $\sigma_{Z}^{2}=0.0625$ and
$R_{Z}=0.909$, the MFVB takes only $11.52$ seconds per replication. If
$51,000$ (resp. $101,000$ and $201,000$) draws are used, the computing time
per replication for the Gibbs sampling of the BSURME model is about $65.64$
(resp. $142.07$ and $352.29$) seconds using a MacBook Pro, 2.8 GHz core i7
with 16Go 1600 MHz DDR3 RAM.\label{footnote_label_1}}. We also compare the
results with the usual frequentist SUR estimation and the standard Bayesian
estimation of SUR model. The Gibbs sampling algorithm for the latter is
presented in Appendix~D of the supplementary material.

\subsection{Case I: Altering $\sigma_{Z}^{2}$}

Amongst the first set of simulation studies labeled Case I,
Table~\ref{Case1:Sim1FreqSUR} presents the results from the frequentist
estimation of SUR\footnote{Without any prior information on the measurement
error, the SUR model for $M$ equations is the following: $y_{i}=X_{i}\beta
+W_{i}\gamma +\varepsilon _{i}$ , $\varepsilon _{i}\sim N\left( 0,\Sigma
_{\varepsilon }\right) $ , $i=1,..,N $ where $W_{i}$ is the covariate with
measurement error.} model for the case $R_{Z}=0.8$ and $\sigma _{Z}^{2}=1$.
Results show that estimates are strongly biased mainly for the intercepts
$\beta _{11}$ and $\beta _{21}$, and for $\gamma _{1}$ and $\gamma _{2}.$ The
relative biases ($\hat{\beta} / \beta - 1$) of the intercepts (resp. the
$\gamma$'s) are $38.1\%$ and $28.8\%$, (resp. $-19.8\%$ and $-19.6\%$). The
$\gamma$'s are strongly under-estimated. On the contrary, slope coefficients
$\beta _{12}$, $\beta _{13}$ and $\beta _{23}$ are less contaminated by the
measurement error and have a lower dispersion of the estimated coefficients
than the intercepts. The relative bias of $\beta _{22}$ ($22.3\%$) is close
(in absolute value) to that of $\gamma$'s. Elements of the
variance-covariance matrix $\Sigma _{\varepsilon }$ are strongly
over-estimated with a relative error of $317\%$ and $314\%$ for the variances
and $4.6\%$ for the covariance $\sigma_{12}$. It leads to a strong
under-evaluation  of the coefficient of correlation $\rho_{\varepsilon_1
\varepsilon_2} =0.126$ far from the true value $\left(0.5\right) $. The
Bayesian estimation of SUR model, presented in Table~G1 of the supplementary
material, give similar results. The posterior means of the coefficients
(resp. posterior standard errors) are similar to the frequentist coefficient
estimates (resp. standard errors) of the SUR model. The $95\% $ highest
posterior density intervals (HPDI) are also close to the 95\% confidence
interval of the frequentist estimation. The estimated correlation coefficient
$\rho_{\varepsilon_1 \varepsilon_2} =0.125$ is similar to the frequentist
estimate. We also report Geweke's convergence diagnostic (CD), which tests
for the equality of means of the first and last part of a Markov chain on the
basis of samples drawn from the stationary distribution of the chain. In more
than $98\%$ of cases, Geweke's CD (under the null hypothesis, $CD \sim
N(0,1)$) accepts the null hypothesis at $5\%$ level, which suggests that a
sufficiently large number of draws has been taken. Moreover, inefficiency
factors (reported in Table G1) are also close to 1, which confirms that the
chain is mixing well.

In the upper panel of Table~G2 (of the supplementary material), we see how
the Bayesian estimation of SURME model improves the results. The intercepts
$\beta_{11}$ and $\beta_{21}$ are now less biased as compared to that of SUR
model in Table~G1. Their relative errors are $-3.5\%$ and $-7.1\%$,
respectively. This is also true for the other slope coefficients $\beta$.
Moreover, the model neatly corrects the measurement errors and results in
better estimates of $\gamma_{1}$ and $\gamma_{2}$, their relative errors
being $2.1\%$ and $2.6\%$, respectively. We also note that the
variance-covariance matrix is precisely estimated leading to a correlation
coefficient $\rho_{\varepsilon_1 \varepsilon_2} =0.494$. The parameters
$\sigma_{Z}^{2}$ and $\sigma_{u}^{2}$ are well estimated with small posterior
standard deviations and small relative errors ($-2.6\%$ and $0.6\%$,
respectively). But inefficiency factors are large indicating strong
autocorrelation in MCMC draws, particularly for $\gamma_1$ and $\gamma_2$
whose inefficiency factors are $8.62$ and $10.52$, respectively. In more than
$90\%$ of cases, the Geweke's CD confirms that a sufficiently large number of
draws has been taken. The improvement obtained with a SURME (as compared to
the SUR) is interesting and emphasizes the need to model measurement error.
The lower panel of Table~G2 (in the supplementary material) presents the
results of the exposure equation from the SURME model. They show that the
biases are negligible, the posterior standard deviations are small and so are
the inefficiency factors.

Overall, the SURME model is well estimated, but the high autocorrelation in
MCMC draws of $\gamma$ needs additional consideration. According to
\citet{Owen-2017}, the problem of high autocorrelation can be dealt with
thinning, which itself can be optimized according to the cost of computing
the quantities of interest (after advancing the Markov chain) and the speed
at which autocorrelations decay. As shown in Table~G3 (see the supplementary
material), autocorrelations between the successive draws of $\gamma_1$ and
$\gamma_2$, denoted $\rho_{\tau} (\gamma_1)$ and $\rho_{\tau} (\gamma_2)$,
are close to one and the rate of decay is very slow. For example, $\rho_{1}
(\gamma_1)= 0.98$, $\rho_{10} (\gamma_1)= 0.82$ and $\rho_{1} (\gamma_2)=
0.98$, $\rho_{10} (\gamma_2)= 0.87$. The autocorrelations of some latent
variables $Z_i$ are slightly higher ($0.995$) than those of the $\gamma$'s,
but not reported for the sake of brevity. The cost of computing of
$\tilde{Z}_i$ is on average $2.71$ and an autocorrelation of $0.995$ leads to
an optimal thinning of factor $k=86$ (see appendix F and Table F1 in the
supplementary material). Henceforth, we use a thinning of factor $k=100$ for
all simulations.

We re-estimate the Bayesian SUR and SURME models with a thinning of 100, but
only report the results for SURME. The results, presented in the upper panel
of Table~\ref{Case1:Sim1SURMEthin}, show that the posterior means and
standard deviations are close (or identical) to those of Table G2 (in the
supplementary material). Values of the inefficiency factors are small and are
all between $(1.004, 1.23)$. Specifically, the reduction in inefficiency
factor is tremendous for $\gamma$'s (\textit{e.g.}, $1.11$ \textit{versus}
$8.62$ for $\gamma_1$ and $1.23$ \textit{versus} $10.52$ for $\gamma_2$). The
lower panel of Table~\ref{Case1:Sim1SURMEthin} presents the results for the
exposure equation in the SURME model. Once again, the results show that the
biases are negligible, the posterior standard deviations are small and the
inefficiency factors and Geweke's CD suggest good mixing of the MCMC draws.
Specifically, the autocorrelations of $\gamma_1$ and $\gamma_2$ are now small
($\rho_{1} (\gamma_1)= 0.15$, $\rho_{1} (\gamma_2)= 0.27$) and quickly
converge towards zero ($\rho_{10} (\gamma_1)= -0.007$, $\rho_{10} (\gamma_2)=
-0.003$) confirming a good mixing of Markov chains (see Table~G4 in the
supplementary material).

\begin{sloppypar}
To compare models, we employ the deviance information criterion or
DIC\footnote{Note that there does not exist any model adequacy measure that
takes into account measurement error in a multi-equation setup. This is an
open area of research and the only related work is \citet{Cheng-etal-2014},
where they propose a coefficient of determination for linear regression
models with measurement error.} proposed by \citet{Spiegelhalter-etal-2002},
and further studied in \citet{Celeux-etal-2006} and
\citet{Spiegelhalter-etal-2014}. Following \citet{Chan-Grant-2016}, we
compute the integrated likelihood for the SUR and SURME model with a thinning
factor of 1 and 100. This is used to calculate the marginal likelihood, which
is then utilized in DIC and the effective number of parameters $p_{D}$. For
the SUR model, we get negative estimates of $p_{D}$ which is indicative of
either a poor fit between the model and data or a conflict between the prior
and data. Different variations on the prior yield negative $p_{D}$, so it is
more likely due to a poor fit between the SUR model and data. When $p_{D}<0$,
the DICs are not adequate for evaluating the complexity and the fit of a
model \citep{Celeux-etal-2006}. On the other hand, for the SURME model we get
a positive estimate of $p_{D}$, synonymous with better fit (see Appendix~E in
the supplementary material for further discussion of the method and Table~G5
for the results).
\end{sloppypar}

We next discuss the results from MFVB approach, which on average takes about
$145$ cycles to get the maximum of the evidence lower bound $l$ and the
algorithm is terminated when the relative increase in the evidence lower
bound $l$ is less than $10^{-7}$. The results from the MFVB estimation of
SURME model are presented in Table~\ref{Case1:Sim1MFVB}, which shows that the
MFVB approach gives better results compared to Gibbs sampling. All parameters
have similar or lower biases, mainly for the intercepts $\beta_{11}$,
$\beta_{21}$ and for $\gamma$. However, the relative biases of the $\gamma$'s
are now reduced ($0.7\%$ and $-0.3\%$) as compared to Bayesian estimation of
SURME model. The estimates for $\sigma_{Z}^{2}$ and $\sigma_{u}^{2}$ show
that the model accurately estimates the variances and their relative biases
are small ($0.4\%$ and $-3.5\%$, respectively). The standard deviation of all
the parameters are smaller compared to those from MCMC estimation leading to
slightly narrower $95\%$ credible intervals (as compared to the $95\%$
HPDI)\footnote{When calibrating this Monte Carlo study, we found a
significant underestimation of the variances of the coefficients $\gamma
_{1}$ and $\gamma_{2}$, echoing the previous discussion around the work of
Blei \textit{et al.} (2017) (Section 3). After several trials (and to avoid
embarking on more complex approaches such as linear response variational
Bayes \citep{Giordano-etal-2018} or $\alpha $-variational inference
\citep{Yang-etal-2018}, we decided to use the following simple trick to
correct this undervaluation: $\sigma_{\gamma_{j}}$ is replaced by
$\sigma_{\gamma_{j}} \times \sqrt{MK/ E_{q\left( \sigma _{Z}^{2} \right) }}$,
for $j=1,..,M$ (see Section~B2 of the supplementary material).}.
Additionally, estimates of $\sigma _{mm^{\prime }}$ are closer to their
theoretical values and the estimated correlation coefficient
$\rho_{\varepsilon_1 \varepsilon_2} =0.488$ is close to 0.5, the actual
value. The lower panel of Table~\ref{Case1:Sim1MFVB} presents the results
from the exposure equation which emphasizes the accurate estimation of the
$\omega$ parameters. The MFVB approximation of both the classical structural
form and the exposure model shows that there are definite advantages in
adopting the MFVB approach to estimate measurement error models as compared
to the pure Bayesian method.

We next decrease the variance $\sigma_{Z}^{2}$ from $\sigma_{Z}^{2}=1$ to
$\sigma_{Z}^{2}=0.0625$ leading to $\sigma_{u}^{2}= 0.0156$. The results are
presented in Tables~G6 to G9 of the supplementary material. Results from the
frequentist and Bayesian estimation of SUR model always reveal strong
over-estimation of the intercepts, $\beta_{22}$ and strong under-estimation
of the slopes $\gamma $ of the error prone covariate $Z_{i}$. However,
over-estimation of the variances $\sigma_{11}$ and $\sigma_{22}$ ($ \simeq
19\%$ for both) are largely reduced as compared to those of
$\sigma_{Z}^{2}=1$, but leads to a slightly under-estimated correlation
coefficient $\rho =0.42$. When we incorporate the measurement error in the
model, \textit{i.e.}, SURME model with a very small variance of
$\sigma_{Z}^{2}$, the Bayesian estimates show a less accurate estimation of
the intercepts (increasing the negative relative biases $-29.4\%$ and
$-25.6\%$), of the $\gamma$'s ($15\%$ and $16.8\%$) and of all the $\beta$'s.
Moreover, inefficiency factors rise to about $2$ indicating a relative loss
of efficiency due to slightly correlated samples. To neutralize this effect,
we can increase the thinning appropriately\footnote{We relaunched the
simulations for this case with a thinning of $120$ and we find inefficiency
factors close to $1$. The results are available upon request for the sake of
brevity.}. Results for the exposure equation in the SURME model do not seem
to be affected by the strong decrease of the variance $\sigma_{Z}^{2}$. The
use of the MFVB approximation significantly attenuates the biases observed
with the Bayesian estimation of SURME. The relative biases for the intercepts
are now $-17\%$ and $-11\%$, and those for the $\gamma$'s are $8.6\%$ and
$7\%$, respectively. The relative errors for the variances $\sigma _{mm},\,
(m =1,2)$ reduce to $-2.5\%$ approximately and we get an estimated
correlation coefficient $\rho =0.52$. The MFVB approximation accurately
estimates parameters of the exposure equation. Once again, the MFVB method
reveals its advantages in estimating a SUR model with measurement error
although this advantage tends to be attenuated when a very small variance
$\sigma_{Z}^{2}$ occurs.

In summary, for a fixed measurement error of $25\%$, increasing the variance
$\sigma _{Z}^{2}$ of the error prone covariate $Z_{i}$ strongly biases the
estimated variances $\sigma _{mm} \, (m=1,2)$ as well as the whole set of
coefficients (intercepts and slopes) in the SUR model irrespective of the
method of estimation. But, taking into account the measurement errors through
SURME model neutralizes the negative effects of the increasing uncertainty on
the error prone covariate $Z_{i}$ and thus strongly reduces, or even
eliminates the biases to obtain satisfactory estimates. This conclusion is
further reinforced with the use of the MFVB approximation.

\subsection{Case II: Altering $R_{Z}$}

We now investigate the performance of the proposed algorithms where
$\sigma_{Z}^{2}=0.0625$ and the reliability ratio $R_{Z}$ is gradually
decreased. Specifically, we consider $R_{Z}=\left\{\, 0.8, \,
0.5714\right\}$, which leads to $\sigma_{u}^{2}=\sigma
_{Z}^{2}(1-R_{Z})/R_{Z} = \left\{0.0156, \, 0.0469\right\} $ and
noise-to-true variance ratio $(1-R_{Z})/R_{Z}$ of $\left\{25\%, \,
75\%\right\}$. In the previous subsection, we have already studied the case
where $\sigma_{Z}^{2}=0.0625$ and $R_{Z}=0.8$, therefore the focus is only on
the case where the reliability is reduced to 0.5714.

The results presented in Tables~G14-G17 of the supplementary material are
poorer than those of $R_{Z}=0.8$ for both the frequentist and Bayesian
estimates of SUR model, with stronger over-estimation of the intercepts
($84.3\%$ and $25.6\%$) and stronger under-estimation of the slopes $\gamma$
($-42.5\%$). The relative biases of the intercepts are larger than in the
previous cases and the same is true for the $\gamma$'s and even more obvious
for the $\sigma_{mm} \, (m=1,2)$ (approximately $42\%$). The estimated
correlation coefficient $\rho =0.38$ is far from the true value $0.5$. The
Bayesian estimates of SURME model show a significant improvement, reducing
the biases for the intercepts ($33\%$ and $-10.8\%$) and the $\gamma$'s
($16.2\%$ and $18.4\%$), but with slightly larger posterior standard
deviations. The variance-covariance matrix $\Sigma_{\varepsilon}$ is well
estimated, with small relative biases of $-7\%$ and $-9\%$ for $\sigma_{mm}\,
(m=1,2)$. The estimated correlation coefficient turns out to be $\rho =0.53$.
Both $\sigma_{Z}^{2}$ and $\sigma_{u}^{2}$ are also close to the true values
(their relative biases are $-15.2\%$ and $15.1\%$, respectively). Once again,
the improvement with the MFVB approximation is more noticeable as we get
better results for the parameters with slightly smaller standard deviations.
The relative biases for the intercepts are $-16.6\%$ and $-5.7\%$, and those
of the $\gamma$'s are $8.5\%$ and $6.5\%$. For the slope coefficients, the
relative biases range between $-11\%$ and $-2.5\%$. Both $\sigma_{Z}^{2}$ and
$\sigma_{u}^{2}$ are also better estimated (their relative biases are
respectively $-5.6\%$ and $4.4\%$). This is also true for the $\sigma_{mm}\,
(m=1,2)$ (with small relative biases of $-1.9\%$ and $-1.6\%$) leading to an
estimated correlation coefficient $\rho =0.51$.

To summarize, a change in the reliability ratio $R_{Z}$ --- for example,
increasing the measurement error from $25\%$ to $75\%$ --- strongly biases
the whole set of coefficients (intercepts and slopes), including the
estimated variances $\sigma_{mm} \, (m=1,2)$ in the SUR model. This is true
both for the frequentist and Bayesian approach. On the other hand, accounting
for measurement error through SURME model largely eliminates the negative
effects of this alteration and strongly reduces the biases in SUR estimation.
Moreover, the use of MFVB approximation improves the results beyond those
obtained with the Bayesian estimation of SURME model.\footnote{To get results
with the Bayesian estimation of the SURME equivalent to those obtained with
the MFVB approximation, it should be necessary to greatly increase the number
of MCMC draws resulting in a very important cost in terms of computing time.
Going from $51,000$ draws to $201,000$ draws leads to a relative increase in
computing time per replication from $16$ to $86$ times that of MFVB (see note
\footref{footnote_label_1}). There is therefore an obvious trade-off against
the Bayesian estimation of the SURME and in favor of the MFVB approximation.}

Finally, we present in Table~\ref{Table:SummaryRelErrors} the relative errors
of parameters for all the cases\footnote{For $\sigma_{Z}^{2}=1$ and
$R_{Z}=0.5714$, results are given in Tables~G13-G18. Last, Table~G25 gives a
summary of DICs and $p_D$s.} of $\sigma_{Z}^{2}=\left\{1, \, 0.0625\right\} $
and $R_{Z}=\left\{0.8, \, 0.5714\right\}$.  At a glance, this table allows us
to compare and contrast all of the previously discussed results and another
case provided in the supplementary material. To reiterate, the results show
that for a fixed measurement error, increasing the variance of the error
prone covariate $Z_{i}$ strongly biases the whole set of coefficients
(intercepts and slopes) as well as the estimated variances $\sigma_{mm} \,
(m=1,2)$ in the SUR model, irrespective of the method of estimation. This is
also the case when, for a fixed variance $\sigma_{Z}^{2}$, the reliability
ratio $R_{Z}$ is reduced. Fortunately, taking into account the measurement
error through SURME model attenuates or even neutralises the undesirable
effects of the increasing uncertainty on the error prone covariate $Z_{i}$ or
reducing the reliability ratio. This conclusion is further reinforced with
the use of the MFVB approximation.

\section{Application}
\begin{sloppypar}
The statistical literature on modeling measurement error has often drawn
applications from health and epidemiology studies where certain variables
such as urinary sodium chloride \citep{Liu-Liang-1992} and blood pressure
\citep{Kannel-etal-1986} are treated as measured with error. In this context,
\citet{Carroll-etal-Book-2006} utilizes measurement error in systolic blood
pressure (\emph{SBP}) on several occasions to illustrate different kinds of
measurement error models and estimation methods. The idea is that long-term
\emph{SBP} is extremely difficult to measure and hence all recorded
observations from clinic visits on \emph{SBP} have measurement error. We draw
motivation from \citet{Carroll-etal-2006} and
\citet{Tao-etal-2011}\footnote{The Association for the Advancement of Medical
Instrumentation (AAMI) and the British Hypertension Society recommend an
absolute mean deviation (between oscillometric and invasive measurements of
systolic blood pressure) less that $5$ mmHg and a standard deviation of less
than $8$ mmHg. Using $6640$ systolic blood pressure measures from $270$
participants, \citet{Tao-etal-2011} find that large measurement errors of $
> 10$ mmHg, (i.e., oscillometric measurement overestimates the real SBP) in
$28.78\%$ of the sample when their SBP values are around $90$ mmHg. They also
find that when SBP is more than $150$ mmHg, most of the measurement errors
are negative (i.e., oscillometric measurement underestimates the real SBP) .
In their study, \citet{Tao-etal-2011} found an absolute mean deviation of
$1.98$ mmHg but a standard deviation of $14.87$ mmHg, so practically doubling
the recommended norm for measurement errors. As the authors say (p.288)
\textquotedblleft\textit{If oscillometric measurement underestimates the real
BP around the critical value (90 mmHg), the physician may give a wrong
treatment. If the oscillometric measurement overestimates the real BP around
90 mmHg, the error may lead to an under-diagnosis and a delayed treatment
response to perioperative hypotension and significantly increases the risk of
dying}''.} and present an application of SURME model where the primary
objective is to model the measurement error in \emph{SBP} and explore the
possibility of a better model fit relative to a standard SUR model.
\end{sloppypar}

The current study utilizes data from the National Health and Nutrition
Examination Survey (NHANES) for 2007-2008, a widely used survey designed to
assess health and nutritional status of civilians, non-institutionalized
adults and children in the United States. NHANES collects data by
interviewing individuals at home, who then report to mobile examination
centers (MECs) to complete the health examination component of the survey.
The MEC's provide a standardized environment for the collection of high
quality data, thus favoring dependable statistical estimation and
interpretations. The survey is unique in the sense that it combines
interviews and physical examinations of the respondents.

The dependent variables in the model are log of weight and high density
lipoprotein ($HDL$), which is also known as `good cholesterol'. The
covariates that are common to both equations and assumed to be measured
without error are as follows: age, gender, smoking status, hours of sedentary
activities, sleep disorder and low density lipoprotein plus 20 percent of
Triglyceride ($LDL20T$). The variable `height' is only expected to effect
weight, not $HDL$, and therefore only included in the log weight equation.
Observed $SBP$ is assumed to have measurement error and transformed as $\ln
(SBP-50)$ to avoid scaling problems, as done in \citet{Carroll-etal-2006}.
The third reading on $SBP$ is used as data and the first two readings are
utilized to form priors on relevant parameters. Focussing on adults and
removing missing observations on all variables of interest leaves us with a
total of $N=1,001$ observations. Table~\ref{Table:AppDescStat} presents the
definition and descriptive statistics of all the variables used in the study.

To estimate the different SUR models with and without measurement error, we
utilize the following relatively vague priors on the parameters: $\beta \sim
N_{15}\left( \beta _{0},B_{0}\right) $ with $\beta _{0}=0_{15}$, $B_{0}=10
I_{15}$, $\gamma \sim N_{2}\left( \gamma _{0},G_{0}\right) $ with $\gamma
_{0}=0_{2}$, $G_{0}=10 I_{2}$, $\Sigma _{\varepsilon }\sim IW_{2}\left( \nu
_{0},S_{0}\right) $ with $\nu _{0}=10$ and $S_{0}=10 I_{2}$, $\omega \sim
N\left( \omega _{0},O_{0} \right) $ with $\omega_{0}=0_{15}$, $O_{0}=I_{15}$,
$\sigma _{Z}^{2}\sim IG\left( 50,10\right) $ and $\sigma _{u}^{2}\sim
IG\left( 50,5\right)$. The prior distribution for $\sigma _{Z}^{2}$ is
specified such that the prior mean ($0.2$) is close to the mean difference
between first and second readings on transformed \emph{SBP} ($0.024$).
Similarly, the prior distribution for measurement error variance $\sigma
_{u}^{2}$ is stipulated such that prior mean ($0.1$) is near the mean
difference in variance from first and second readings on transformed
\emph{SBP} ($0.002$). Note that some of the parameters only appear in the
measurement error model and the priors are used accordingly.

We first look at the results for the Bayesian estimation of SUR model
presented in Table~\ref{Table:AppBSUR} from 400,000 draws after a burn-in of
50,000 draws with a thinning factor of 100 (optimized following the approach
in \citet{Owen-2017}). The posterior estimates show that $\ln (age)$ is not
statistically different from zero in both the $\ln (weight)$ and $HDL$
equations. \emph{Male} indicator variable positively affects $\ln (weight)$,
but negatively affects \emph{HDL}. \emph{Height} has a strong positive effect
on $\ln (weight)$ and this is typically anticipated for all adults. Smoking
daily or some days is negatively associated with $\ln (weight)$. This outcome
is not surprising since smoking is well known to reduce appetite. On the
contrary, smoking seems to have no significant effect on $HDL$. Number of
hours of sedentary activities is positively associated with $\ln (weight)$
and negatively associated with \emph{HDL}. The result confirms the generally
held belief that being inactive increases weight and is negatively associated
with good cholesterol. Sleep disorder is also known to be positively
associated with weight gain and this is confirmed in our findings but it has
no effect on \emph{HDL}. \emph{LDL20T} has a positive (negative) effect on
$\ln (weight)$ (\emph{HDL}), which is expected since \emph{LDL} is commonly
referred as `bad cholesterol' and is associated with weight gain. Transformed
\emph{SBP} has a positive effect on $\ln (weight)$, but a negative effect on
\emph{HDL} which is statistically different from zero at 90\% probability
level. As the first equation is a log-log specification, the coefficient of
$\ln(SBP-50)$ is an elasticity. Then, a $10\%$ increase in the transformed
\emph{SBP} leads to a $0.967\%$ growth of weight. The second equation is a
semi-log specification, so the elasticity of \emph{HDL} relative to the
transformed \emph{SBP} at the mean of the sample is: $0.1012/1.32 = 0.076$. A
$10\%$ increase in the transformed $SBP$ leads to a $0.76\%$ growth of $HDL$.
The estimated correlation coefficients of the residuals between the two
equations is $\rho_{\varepsilon_1 \varepsilon_2} =-0.304$. Inefficiency
factors are close to 1 suggesting a good mix of the draws and the Geweke's CD
($CD \sim N(0,1)$) confirms that a sufficiently large number of draws has
been taken.

\begin{figure}[!t]
\centering
\makebox[\textwidth]{%
\includegraphics[width=6in,height=4in]{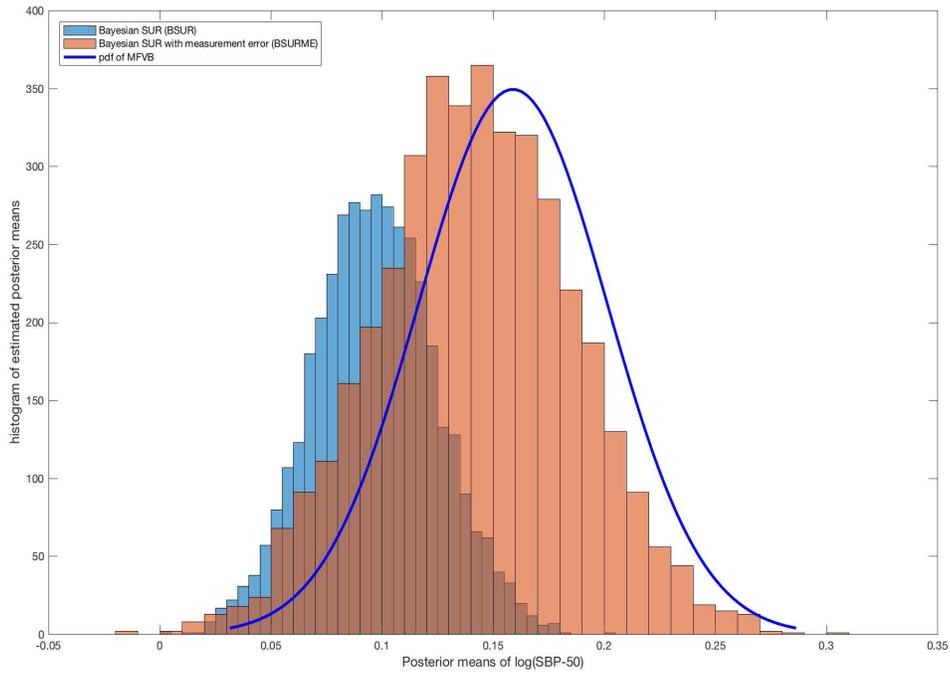}%
}%
\caption{Posterior means of $\log$(SBP-50) for $\log$(weight) equation.}
\label{Fig:sbpWEIGHTeq}
\end{figure}

The results from the Bayesian estimation of SURME model (which accounts for
the measurement error in the covariate \emph{SBP}) is presented in the upper
panel of Table~\ref{Table:AppBSURME}. A quick glance shows that the results
for the covariates measured without error are similar to those in
Table~\ref{Table:AppBSUR}, except for the intercept and the male indicator in
the $\ln (weight)$ equation. The $\ln (height)$ coefficient in the $\ln
(weight)$ equation is now slightly higher but its $ 95\%$ credible interval
overlaps with that of the SUR model. Posterior estimates corresponding to
transformed $SBP$ increase in both equations (from $0.097$ to $0.141$ in the
$\ln(weight)$ equation and from $0.101$ to $0.152$ in the $HDL$ equation)
leading to the following elasticities at the mean of the sample: $0.141$ and
$0.115 (= 0.152/1.32)$, for weight and $HDL$, respectively. However, as the
posterior standard errors become larger (from $0.027$ to $0.044$ in the
$\ln(weight)$ equation and from $0.056$ to $0.092$ in the $HDL$
equation)---as in the Monte Carlo study---the $95\%$ HPDI of the posterior
means of $\ln(SBP-50)$ overlap even if the distribution moves to the right
when we go from SUR to SURME (see Figure~\ref{Fig:sbpWEIGHTeq}). In
particular, the $95\%$ HPDI of the posterior means of $\ln(SBP-50)$ are
$\left[ 0.051;0.143\right] $ and $\left[ 0.009;0.194\right] $ in the $\ln
(weight)$ and the $HDL$ equations, respectively, in the SUR model, and
$\left[ 0.067;0.213\right]$ and $\left[ 0.005;0.304\right] $ in the $\ln
(weight)$ and the $HDL$ equations in the SURME model. In the $HDL$ equation,
the coefficient of \emph{SBP} is positive but statistically equivalent to
zero. Posterior estimate of measurement error variance is $0.029$, which
leads to an estimated reliability ratio of about $ 59.78\%$ and a
noise-to-true variance ratio of $67.27\%$. The posterior variances of the
disturbances from $\ln (weight)$ and $HDL$ are close to those of SUR model
and lead to an error correlation $\rho_{\varepsilon_1 \varepsilon_2} =-0.25$.
Inefficiency factors are close to 1 and Geweke's CD confirms, for most
parameters, that a sufficiently large number of draws has been taken.

The lower panel of Table~\ref{Table:AppBSURME} presents the results for the
exposure equation from the Bayesian estimation of SURME model. In the first
equation, only three variables have a positive effect on $\ln(SBP-50)$:
$\ln(age)$, LDL20T and $\ln (height)$ (only at the $10\%$ probability level).
In the second equation, four variables have a positive effect on
$\ln(SBP-50)$: $\ln(age)$, LDL20T, male and smokers (the last two variables
are different from zero only at the $10\%$ probability level).

We next estimate the SURME model using the MFVB approach, which takes $1565$
cycles to get the maximum of the evidence lower bound $l$. The results,
presented in the upper panel of Table~\ref{Table:AppMFVB}, shows that for the
transformed \emph{SBP} both the coefficient ($0.159$) and the probability
interval ($\left[ 0.076;0.242\right]$) are similar to those obtained from the
Bayesian estimation of SURME model. In the $HDL$ equation, the marginal
effect at $0.186$ is higher compared to the Bayesian estimate, with a $95\%$
probability interval of $\left[0.03;0.34\right]$. Taking into account
measurement errors using MFVB allows to get significantly larger and more
accurate elasticities of weight ($0.159$) and $HDL$ ($0.1414= (0.186/1.32)$)
for the transformed $SBP$ compared to the other method.

Posterior estimate of measurement error variance from the MFVB approach is
$0.029$, which is similar to the Bayesian estimate, and leads to an estimated
reliability ratio of about $60.2\%$ and to a noise-to-true variance ratio of
$66.1\%$. The posterior variances of the disturbances from $\ln (weight)$ and
$HDL$ are close to the Bayesian estimates and lead to the same correlation
between the errors of the two equations $\rho_{\varepsilon_1 \varepsilon_2}
=-0.25$. The lower panel of Table~\ref{Table:AppMFVB} presents results for
the exposure model estimated using with MFVB approximation method. In the
first equation, four variables have a positive effect on $\ln(SBP-50)$:
$\ln(age)$, smokers, $LDL20T$ and $\ln (height)$. Similarly, in the second
equation four variables have a positive effect on $\ln(SBP-50)$: $\ln(age)$,
$male$, $smokers$ and $LDL20T$. The exposure equation in the SURME model
allows to define the implicit links between the true systolic blood pressure
and the ``risk factors'' such as age, gender, smokers and ``bad cholesterol''
($LDL20T$).

\begin{figure}[!b]
\centering
\makebox[\textwidth]{%
\includegraphics[width=6in,height=4in]{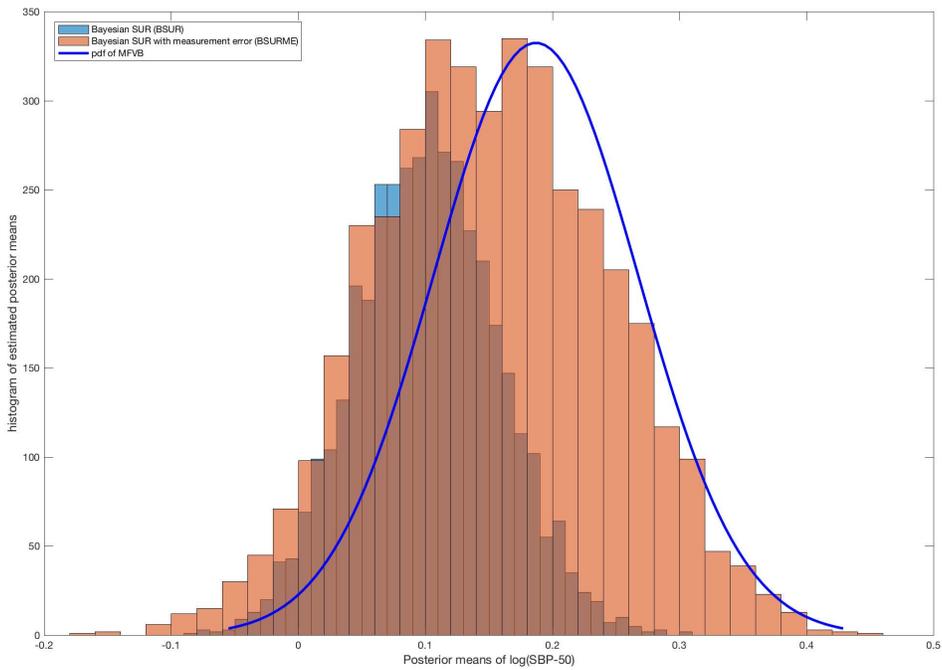}
}%
\caption{Posterior means of $\log$(SBP-50) for $\log$(HDL) equation.}
\label{Fig:sbpHDLeq}
\end{figure}

Figure~\ref{Fig:sbpWEIGHTeq} gives the posterior densities of the parameter
corresponding to $\ln(SBP-50)$ in the $\ln(weight)$ equation from the
Bayesian estimation of SUR model, SURME model and the MFVB estimation of
SURME model. We note a shift of the marginal effect of $\ln(SBP-50)$ on
$\ln(weight)$ to the right of the distribution from a mode established around
$0.097$ for SUR model to a mode of $0.141$ for SURME model but with a wider
dispersion. The estimated probability density function (\emph{pdf}) with MFVB
is slightly to the right and centered around the mode ($0.159$) but with a
surface under the curve globally equivalent to that of Bayesian estimation of
SURME model. In Figure~\ref{Fig:sbpHDLeq}, we observe similar shifts in the
posterior density of the parameter corresponding to $\ln(SBP-50)$ in the
$HDL$ equation, when we move from SUR ($0.101$) to SURME ($0.152$) or to MFVB
($0.187$) estimation of SURME model.

We note that similar to the Monte Carlo study, the MFVB approach to
estimating the SURME model improves the results compared to those from Gibbs
sampling. The results actually lend credibility to the proposed MFVB
algorithm since coefficient estimates for variables which do not have
measurement error are almost unaltered. However, when measurement error in
$SBP$ is ignored as in the SUR model, the posterior estimates are
underestimated relative to the MFVB estimates. So, accounting for measurement
error potentially corrects or reduces the bias in parameter estimates
\citep[see][]{Carroll-etal-2006}.

\section{Conclusion}

The paper considers a SURME model (seemingly unrelated regression where some
covariates have classical measurement error of the structural form) and
introduces two novel estimation methods: a pure Bayesian algorithm based on
MCMC and a second algorithm based on mean field variational Bayes
approximation. The proposed algorithms use a prior distribution on
measurement error variance to resolve identification issues in the model. In
the MCMC estimation, Gibbs sampling is employed to sample the parameters from
the conditional posterior distributions. While most of the conditional
posterior densities have the standard form and are easily derived, the
conditional posterior density for the true unobserved quantity associated
with covariates having measurement error requires extensive attention to
arrive at a manageable form. We also note that the proposed SURME model as
explained is based on the structural form of measurement error, but the
functional form of measurement error can be easily incorporated by
introducing the distribution of the true unobserved quantity as a part of the
subjective prior information. The expression for the joint and conditional
posteriors will remain unchanged. However, estimating the SURME model using
MCMC leads to high autocorrelation in the draws corresponding to the
covariate measured with error. While this is easily dealt using
\emph{thinning}, the paper also proposes the MFVB approach as an alternative
to get around the problem of high autocorrelation.

The proposed estimation algorithms are illustrated in multiple Monte Carlo
simulation studies. While, the first set of 2 simulations (labeled Case~I)
investigate the effect on estimates by varying the variance of the true
unobserved variable (for a fixed reliability ratio), the second set of 2
simulations examine the effect for a changing reliability ratio (for a fixed
variance of the true unobserved variable). The results from all the
simulations show that the Bayesian and MFVB estimation of SURME model reduce
the biases to obtain satisfactory estimates as compared to estimates from SUR
model. Moreover, the MFVB approach turns out as an excellent alternative to
the MCMC and its poor mixing properties in the presence of latent variables.
Besides, the MFVB approach has slightly better estimation accuracy and can be
advantageous with large data sets.

The proposed models and techniques are also implemented in a health study
where the two dependent variables, log of weight and high density lipoprotein
(\emph{HDL}), are regressed on a set of covariates measured without error and
on systolic blood pressure (\emph{SBP}) known to have measurement error. The
model is estimated using the two algorithms and the results obtained reveal
that the sign of the estimated coefficients are mostly consistent with what
is typically found in the literature. Specifically, \emph{SBP} has a positive
effect on both $\ln(weight)$ and \emph{HDL}, measurement error variance is
small with an estimated reliability ratio of about $60\%$ and a noise-to-true
variance ratio of $66\%$. To offer a baseline comparison, a SUR model that
ignores measurement error in \emph{SBP} is also estimated using Gibbs
sampling. Comparing the results across models, we see that posterior
estimates for covariates without measurement error are almost identical, but
that of \emph{SBP} is lower and hence underestimated both in the weight and
the $HDL$ equations.

The combination of SUR and measurement error models is attractive and the
proposed model can be generalized in several directions. One straightforward
extension is the introduction of multiple covariates with measurement error
in each SUR equation. However, the challenge here is to keep track of
measurement errors arising from different covariates. The proposed SURME
model can also be modified by introducing classical measurement error in the
response variable or nonclassical measurement error models, where the errors
may be correlated with the latent true values. Beyond the SUR models, these
Bayesian approaches may be useful for measurement error in simultaneous
equation models. We leave these possibilities for future research.

\clearpage \pagebreak
\pdfbookmark[1]{References}{unnumbered}       

\bibliography{BibSURME}

\begin{thebibliography}{}
\newcommand{\enquote}[1]{``#1''}

\bibitem[Ando and Zellner(2010)Ando and Zellner]{Ando-Zellner-2010}
Ando, T. and Zellner, A. (2010), \enquote{Hierarchical Bayesian Analysis of the
  Seemingly Unrelated Regression and Simultaneous Equation Models using a
  Combination of Direct Monte Carlo and Importance Sampling Techniques,}
  \emph{Bayesian Analysis}, 5, 65--95.

\bibitem[Bishop(2006)Bishop]{Bishop-2006}
Bishop, C.~M. (2006), \emph{Pattern Recognition and Machine Learning},
  Springer, New York.

\bibitem[Blei et~al.(2017)Blei, Kuckelbir, and McAuliffe]{Blei-etal-2017}
Blei, D.~M., Kuckelbir, A., and McAuliffe, J.~D. (2017), \enquote{Variational
  Inference: A Review for Statisticians,} \emph{{Journal of the American
  Statistical Association}}, 112, 859--877.

\bibitem[Carroll et~al.(2006a)Carroll, Ruppert, Stefanski, and
  Crainiceanu]{Carroll-etal-Book-2006}
Carroll, R.~J., Ruppert, D., Stefanski, L.~A., and Crainiceanu, C.~M. (2006a),
  \emph{Measurement Error in Nonlinear Models: A Modern Perspective}, Chapman
  \& Hall, Boca Raton.

\bibitem[Carroll et~al.(2006b)Carroll, Midthune, Freedman, and
  Kipnis]{Carroll-etal-2006}
Carroll, R.~J., Midthune, D., Freedman, L.~S., and Kipnis, V. (2006b),
  \enquote{Seemingly Unrelated Measurement Error Models with Application to
  Nutritional Epidemiology,} \emph{Biometrics}, 62, 75--84.

\bibitem[Casella and George(1992)Casella and George]{Casella-George-1992}
Casella, G. and George, E.~I. (1992), \enquote{Explaining the Gibbs Sampler,}
  \emph{The American Statistician}, 46, 167--174.

\bibitem[Celeux et~al.(2006)Celeux, Forbes, Robert, and
  Titterington]{Celeux-etal-2006}
Celeux, G., Forbes, F., Robert, C.~P., and Titterington, D.~M. (2006),
  \enquote{Deviance Information Criteria for Missing Data Models,}
  \emph{Bayesian Analysis}, 1, 651--674.

\bibitem[Chan and Grant(2016)Chan and Grant]{Chan-Grant-2016}
Chan, J. and Grant, A.~D. (2016), \enquote{Fast Computation of the Deviance
  Information Criterion for Latent Variable Models,} \emph{Computational
  Statistics and Data Analysis}, 100, 847--859.

\bibitem[Cheng and {Van Ness}(1999)Cheng and {Van Ness}]{Cheng-VanNess-1999}
Cheng, C.-L. and {Van Ness}, J.~W. (1999), \emph{Statistical Regression with
  Measurement Error}, Arnold Publishers, London.

\bibitem[Cheng et~al.(2014)Cheng, Shalabh, and Garg]{Cheng-etal-2014}
Cheng, C.-L., Shalabh, and Garg, G. (2014), \enquote{Coefficient of
  Determination for Multiple Measurement Error Models,} \emph{Journal of
  Multivariate Analysis}, 126, 137--152.

\bibitem[Chib and Greenberg(1995)Chib and Greenberg]{Chib-Greenberg-1995}
Chib, S. and Greenberg, E. (1995), \enquote{Hierarchical Analysis of SUR Models
  with Extension to Correlated Serial Errors and Time-Varying Parameter
  Models,} \emph{Journal of Econometrics}, 68, 339--360.

\bibitem[Faes et~al.(2011)Faes, Ormerod, and Wand]{Faes-etal-2011}
Faes, C., Ormerod, J.~T., and Wand, M.~P. (2011), \enquote{Variational Bayesian
  Inference for Parametric and Nonparametric Regression With Missing Data,}
  \emph{{Journal of the American Statistical Association}}, 106, 959--971.

\bibitem[Fiebig(2001)Fiebig]{Fiebig-2001}
Fiebig, D.~G. (2001), \enquote{Seemingly Unrelated Regression,} in \emph{A
  Companion to Theoretical Econometrics}, ed. B.~H. Baltagi, pp. 101--121,
  Blackwell Publishing, Massachusett.

\bibitem[Fuller(1987)Fuller]{Fuller-1987}
Fuller, W.~A. (1987), \emph{Measurement Error Models}, John Wiley \& Sons, New
  York.

\bibitem[Geman and Geman(1984)Geman and Geman]{Geman-Geman-1984}
Geman, S. and Geman, D. (1984), \enquote{Stochastic Relaxation, Gibbs
  Distributions, and the Bayesian Restoration of Images,} \emph{IEEE
  Transactions on Pattern Analysis and Machine Intelligence}, 6, 721--741.

\bibitem[Geyer(1991)Geyer]{Geyer-1991}
Geyer, C.~J. (1991), \enquote{Markov chain Monte Carlo Maximum Likelihood,} in
  \emph{Computing Science and Statistics: Proceedings of the 23rd Symposium on
  the Interface}, ed. E.~M. Kemramides, pp. 156--163, Interface Foundation of
  North America, Fairfax Station, VA, USA.

\bibitem[Giordano et~al.(2018)Giordano, Broderick, and
  Jordan]{Giordano-etal-2018}
Giordano, R., Broderick, T., and Jordan, M.~I. (2018), \enquote{Covariances,
  Robustness and Variational Bayes,} \emph{Journal of Machine Learning
  Research}, 19, 1--49.

\bibitem[Griffiths and Chotikapanich(1997)Griffiths and
  Chotikapanich]{Griffiths-Chotikapanich-1997}
Griffiths, W.~E. and Chotikapanich, D. (1997), \enquote{Bayesian Methodology
  for Imposing Inequality Constraints on a Linear Expenditure System with
  Demographic Factors,} \emph{Australian Economic Papers}, 36, 321--341.

\bibitem[Griffiths and Valenzuela(2006)Griffiths and
  Valenzuela]{Griffiths-Valenzuela-2006}
Griffiths, W.~E. and Valenzuela, M.~R. (2006), \enquote{Gibbs Samplers for a
  set of Seemingly Unrelated Regressions,} \emph{Australian \& New Zealand
  Journal of Statistics}, 48, 335--351.

\bibitem[Hu and Wansbeek(2017)Hu and Wansbeek]{Hu-Wansbeek-2017}
Hu, Y. and Wansbeek, T. (2017), \enquote{Measurement Error Models: Editor's
  Introduction,} \emph{Journal of Econometrics}, 200, 151--153.

\bibitem[Jeliazkov(2013)Jeliazkov]{Jeliazkov-2013}
Jeliazkov, I. (2013), \enquote{Nonparametric Vector Autoregressions:
  Specification, Estimation and Inference,} \emph{Advances in Econometrics},
  32, 327--359.

\bibitem[Kannel et~al.(1986)Kannel, Neaton, Wentworth, Thomas, Stamler, Hulley,
  and Kjelsberg]{Kannel-etal-1986}
Kannel, W.~B., Neaton, J.~D., Wentworth, D., Thomas, H.~E., Stamler, J.,
  Hulley, S.~B., and Kjelsberg, M.~O. (1986), \enquote{Overall and Coronary
  Heart Disease Mortality Rates in Relation to Major Risk Factors in 325,348
  men Screened for the MRFIT,} \emph{American Heart Journal}, 112, 825--836.

\bibitem[Koop et~al.(2005)Koop, Poirier, and Tobias]{Koop-etal-2005}
Koop, G., Poirier, D.~J., and Tobias, J. (2005), \enquote{Semiparametric
  Bayesian Inference in Multiple Equation Models,} \emph{Journal of Applied
  Econometrics}, 20, 723--747.

\bibitem[Lee and Wand(2016)Lee and Wand]{Lee-Wand-2016}
Lee, C. Y.~Y. and Wand, M.~P. (2016), \enquote{Streamlined Mean Field
  Variational Bayesian Inference in Multiple Equation Models,}
  \emph{Biometrical Journal}, 58, 868--895.

\bibitem[Link and Eaton(2012)Link and Eaton]{Link-Eaton-2012}
Link, W.~A. and Eaton, M.~J. (2012), \enquote{On Thinning of Chains in MCMC,}
  \emph{Methods in Ecology and Evolution}, 3, 112--115.

\bibitem[Liu(1994)Liu]{Liu-1994}
Liu, J.~S. (1994), \enquote{The Collapsed Gibbs Sampler in Bayesian
  Computations with Applications to a Gene Regulation Problem,} \emph{{Journal
  of the American Statistical Association}}, 89, 958--966.

\bibitem[Liu and Liang(1992)Liu and Liang]{Liu-Liang-1992}
Liu, X. and Liang, K.-Y. (1992), \enquote{Efficacy of Repeated Measures in
  Regression Models with Meausrement Error,} \emph{Biometrics}, 48, 645--654.

\bibitem[MacEachern and Berliner(1994)MacEachern and
  Berliner]{MacEachern-Berliner-1994}
MacEachern, S.~N. and Berliner, L.~M. (1994), \enquote{Subsampling the Gibbs
  Sampler,} \emph{The American Statistician}, 48, 188--190.

\bibitem[Ormerod and Wand(2010)Ormerod and Wand]{Ormerod-Wand-2010}
Ormerod, J.~T. and Wand, M.~P. (2010), \enquote{Explaining Variational
  Approximations,} \emph{The American Statistician}, 64, 140--153.

\bibitem[Owen(2017)Owen]{Owen-2017}
Owen, A.~B. (2017), \enquote{Statistically Efficient Thinning of a Markov Chain
  Sampler,} \emph{Journal of Computational and Graphical Statistics}, 26,
  738--744.

\bibitem[Percy(1992)Percy]{Percy-1992}
Percy, D.~F. (1992), \enquote{Prediction for Seemingly Unrelated Regressions,}
  \emph{{Journal of the Royal Statistical Society -- Series B}}, 54, 243--252.

\bibitem[Pham et~al.(2013)Pham, Ormerod, and Wand]{Pham-etal-2013}
Pham, T.~H., Ormerod, J.~T., and Wand, M.~P. (2013), \enquote{Mean Field
  Variational Bayesian Inference for Nonparametric Regression with Measurement
  Error,} \emph{Computational Statistics and Data Analysis}, 68, 375--387.

\bibitem[Rao et~al.(2008)Rao, Toutenburg, Shalabh, and Heumann]{Rao-etal-2008}
Rao, C.~R., Toutenburg, H., Shalabh, and Heumann, C. (2008), \emph{Linear
  Models and Generalizations: Least Squares and Alternatives}, Springer,
  Berlin.

\bibitem[Shalabh(2003)Shalabh]{Shalabh-2003}
Shalabh (2003), \enquote{Consistent Estimation of Coefficients in Measurement
  Error Models with Replicated Observations,} \emph{Journal of Multivariate
  Analysis}, 86, 227--241.

\bibitem[Spiegelhalter et~al.(2002)Spiegelhalter, Best, Carlin, and van~der
  {L}inde]{Spiegelhalter-etal-2002}
Spiegelhalter, D.~J., Best, N.~G., Carlin, B.~P., and van~der {L}inde, A.
  (2002), \enquote{Bayesian Measures of Model Complexity and Fit,}
  \emph{{Journal of the Royal Statistical Society -- Series B}}, 64, 583--639.

\bibitem[Spiegelhalter et~al.(2014)Spiegelhalter, Best, Carlin, and van~der
  {L}inde]{Spiegelhalter-etal-2014}
Spiegelhalter, D.~J., Best, N.~G., Carlin, B.~P., and van~der {L}inde, A.
  (2014), \enquote{The Deviance Information Criterion: 12 years on,}
  \emph{{Journal of the Royal Statistical Society -- Series B}}, 76, 485--493.

\bibitem[Srivastava and Dwivedi(1979)Srivastava and
  Dwivedi]{Srivastava-Dwivedi-1979}
Srivastava, V.~K. and Dwivedi, T.~D. (1979), \enquote{Estimation of Seemingly
  Unrelated Regression Equations,} \emph{Journal of Econometrics}, 10, 15--32.

\bibitem[Srivastava and Giles(1987)Srivastava and Giles]{Srivastava-Giles-1987}
Srivastava, V.~K. and Giles, D. E.~A. (1987), \emph{Seemingly Unrelated
  Regression Equations Models: Estimation and Inference}, Marcel Dekker, New
  York.

\bibitem[Steel(1992)Steel]{Steel-1992}
Steel, M. F.~J. (1992), \enquote{Posterior Analysis of Restricted Seemingly
  Unrelated Regression Equation Models: A Recursive Analytical Approach,}
  \emph{Econometric Reviews}, 11, 129--142.

\bibitem[Tao et~al.(2011)Tao, Chen, Wen, and Bi]{Tao-etal-2011}
Tao, G., Chen, Y., Wen, C., and Bi, M. (2011), \enquote{Statistical Analysis of
  Blood Pressure Measurement Errors by Oscillometry during Surgical
  Operations,} \emph{Blood Pressure Monitoring}, 16, 285--290.

\bibitem[van Dyk and Park(2008)van Dyk and Park]{vanDyk-Park-2008}
van Dyk, D.~A. and Park, T. (2008), \enquote{Partially Collapsed Gibbs
  Samplers: Theory and Methods,} \emph{{Journal of the American Statistical
  Association}}, 103, 790--796.

\bibitem[Wansbeek and Meijer(2000)Wansbeek and Meijer]{Wansbeek-Meijer-2000}
Wansbeek, T. and Meijer, E. (2000), \emph{Meausrement Error and Latent
  Variables in Econometrics}, Noth Holland, Amsterdam.

\bibitem[Yang et~al.(2018)Yang, Pati, and Bhattacharya]{Yang-etal-2018}
Yang, Y., Pati, D., and Bhattacharya, A. (2018), \enquote{$\alpha$ Variational
  Inference with Statistical Guarantees,}
  \emph{https://arxiv.org/abs/1710.03266}.

\bibitem[Zellner(1962)Zellner]{Zellner-1962}
Zellner, A. (1962), \enquote{An Efficient Method of Estimating Seemingly
  Unrelated Regression and Tests for Aggregation Bias,} \emph{{Journal of the
  American Statistical Association}}, 57, 348--368.

\bibitem[Zellner(1971)Zellner]{Zellner-1971}
Zellner, A. (1971), \emph{An Introduction to Bayesian Inference in
  Econometrics}, John Wiley \& Sons, New York.

\bibitem[Zellner and Ando(2010)Zellner and Ando]{Zellner-Ando-2010}
Zellner, A. and Ando, T. (2010), \enquote{A Direct Monte Carlo Approach for
  Bayesian Analysis of the Seemingly Unrelated Regression Model,} \emph{Journal
  of Econometrics}, 159, 33--45.

\end{thebibliography}
\bibliographystyle{jasa}

\newpage


\begin{table}
\centering \small \setlength{\tabcolsep}{4pt}
\setlength{\extrarowheight}{1.5pt} \caption{Frequentist estimates of SUR Model.
The table presents the true parameter values (\textsc{true}), regression
coefficients (\textsc{coef}), relative error (\textsc{re}), standard error
(\textsc{se}) and 95\% confidence interval (\textsc{inf, sup}). $N=300$,
$\sigma^{2}_{Z} =1$, $R_{Z}=0.8$, Replications=$100$.}
\begin{tabular}{lrrrrrrrrrrr}
\toprule
 & $\beta_{11}$  &  $\beta_{12}$  &  $\beta_{13}$  &  $\beta_{21}$  & $\beta_{22}$
 & $\beta_{23}$  &  $\gamma_1$    &  $\gamma_2$    &  $\sigma_{11}$ & $\sigma_{12}$
 &$\sigma_{22}$  \\ \midrule
\textsc{true}
    &  3  &  5  &  4    &  4  &  3.8  &  3  &  4
    &  4  &  1  &  0.5  &  1  \\ \midrule
\textsc{coef}  &  4.144  &  5.622  &  4.246  &  5.150  &  4.649  &  3.353  &
                  3.206  &  3.216  &  4.171  &  0.523  &  4.144 \\
\textsc{re}    &  0.381  &  0.124  &  0.062  &  0.288  &  0.223  &  0.118  &
                 -0.198  & -0.196  &  3.171  &  0.046  &  3.144\\
\textsc{se}    &  0.350  &  0.219  &  0.106  &  0.348  &  0.233  &  0.112  &
                  0.105  &  0.105  &&&\\
\textsc{inf}   &  3.458  &  5.192  &  4.038  &  4.467  &  4.192  &  3.134  &
                  3.000  &  3.010  &&&\\
\textsc{sup}   &  4.830  &  6.051  &  4.455  &  5.833  &  5.105  &  3.572  &
                  3.412  &  3.422  &&&\\
\bottomrule
\end{tabular}
\label{Case1:Sim1FreqSUR}
\end{table}

\begin{table}
\centering \small \setlength{\tabcolsep}{4pt} \setlength{\extrarowheight}{1.5pt}
\caption{Bayesian estimation of SURME model. The upper panel presents results
for the main equations and the lower panel presents results for the
exposure equations. \textsc{mean} is posterior mean, \textsc{re} is relative error,
\textsc{std} is posterior standard deviation, \textsc{if} is inefficiency factor,
and (\textsc{inf-hpdi, sup-hpdi}) represents 95\% HPDI, and \textsc{cd}
represents Geweke's convergence diagnostics.
$N=300$,  $\sigma^{2}_Z = 1$, $R_Z=0.8$, Draws=$51,000$,
Burnin draws=$1,000$, thinning=$100$, Replications=$100$.}
\begin{tabular}{lrrrr rrrrr rrrr}
\toprule
 &  $\beta_{11}$  &  $\beta_{12}$  &  $\beta_{13}$  &  $\beta_{21}$   &  $\beta_{22}$
 &  $\beta_{23}$  &  $\gamma_1$    &  $\gamma_2$    &  $\sigma^{2}_Z$ &  $\sigma^{2}_u$
 &  $\sigma_{11}$ &  $\sigma_{12}$ &  $\sigma_{22}$  \\ \midrule
\textsc{true}
         &  3   &  5  &  4     &  4  &  3.8  &  3  &  4
         &  4   &  1  &  0.25  &  1  &  0.5  &  1        \\ \midrule
\textsc{mean}  &  2.895  &  4.858  &  3.986  &  3.716  &  3.694  &  2.999
               &  4.083  &  4.103  &  0.974  &  0.251  &  1.025  &  0.504  &  1.015\\
\textsc{re}    & -0.035  & -0.028  & -0.004  & -0.071  & -0.028  & -0.000
               &  0.021  &  0.026  & -0.026  &  0.005  &  0.025  &  0.008  &  0.015\\
\textsc{std}   &  0.369  &  0.243  &  0.119  &  0.376  &  0.261  &  0.125
               &  0.131  &  0.130  &  0.071  &  0.020  &  0.187  &  0.135  &  0.184\\
\textsc{if}    &  1.029  &  1.028  &  1.023  &  1.055  &  1.066  &  1.050
               &  1.117  &  1.232  &  1.032  &  1.011  &  1.011  &  1.004  &  1.013\\
\textsc{inf-hpdi}
               &  2.280  &  4.450  &  3.786  &  3.086  &  3.257  &  2.790
               &  3.870  &  3.893  &  0.862  &  0.220  &  0.752  &  0.297  &  0.746\\
\textsc{sup-hpdi}
               &  3.494  &  5.251  &  4.180  &  4.320  &  4.114  &  3.202
               &  4.301  &  4.320  &  1.093  &  0.284  &  1.359  &  0.738  &  1.347\\
\textsc{cd}    &  0.990  &  0.980  &  1.000  &  0.970  &  0.950  &  0.980
               &  0.990  &  0.950  &  0.960  &  0.990  &  0.990  &  0.960  &  0.970\\
\midrule
\end{tabular}
\begin{tabular}{lrrrr rrr}
& $\omega_{11}$ & $\omega_{12}$ & $\omega_{13}$ & $\omega_{21}$
& $\omega_{22}$  & $\omega_{23}$ &  \\ \midrule

\textsc{true}  &  1.5    &  0.75   &  0.3    &  1.5    &  1.05   &  0.45  & \\
\midrule
\textsc{mean}  &  1.466  &  0.773  &  0.307  &  1.489  &  1.068  &  0.443 & \\
\textsc{re}    & -0.023  &  0.031  &  0.022  & -0.007  &  0.017  & -0.016 & \\
\textsc{std}   &  0.165  &  0.109  &  0.055  &  0.166  &  0.109  &  0.055 & \\
\textsc{if}    &  1.000  &  1.001  &  1.001  &  1.001  &  1.004  &  1.000 & \\
\textsc{inf-hpdi}
               &  1.194  &  0.593  &  0.216  &  1.214  &  0.888  &  0.353 & \\
\textsc{sup-hpdi}
               &  1.735  &  0.952  &  0.396  &  1.758  &  1.246  &  0.533 & \\
\textsc{cd}    &  0.990  &  0.990  &  0.990  &  0.970  &  0.980  &  0.980 & \\
\bottomrule
\end{tabular}
\label{Case1:Sim1SURMEthin}
\end{table}

\begin{table}
\centering \small \setlength{\tabcolsep}{4pt} \setlength{\extrarowheight}{1.5pt}
\caption{MFVB estimation of SURME model. The upper panel presents results for the main
equations and the lower panel presents results for the exposure equations.
\textsc{mean} is posterior mean, \textsc{re} is relative error, \textsc{std}
is posterior standard deviation, and (\textsc{inf-cri, sup-cri}) represents
95\% credible interval. $N=300$,  $\sigma^{2}_Z =1$, $R_Z=0.8$,
Replications=$100$, Cycles = $145.49$, Maximum Elbo = $-3517.205$.}
\begin{tabular}{lrrrrrrrrrrrrr}
\toprule
 &  $\beta_{11}$  &  $\beta_{12}$  &  $\beta_{13}$  &  $\beta_{21}$   &  $\beta_{22}$
 &  $\beta_{23}$  &  $\gamma_1$    &  $\gamma_2$    &  $\sigma^{2}_Z$ &  $\sigma^{2}_u$
 &  $\sigma_{11}$ &  $\sigma_{12}$ &  $\sigma_{22}$  \\ \midrule
\textsc{true}
&  3   &  5  &  4  &  4  &  3.8  &  3  &  4  &  4  &  1  & 0.25  &  1  &  0.5 &  1 \\
\midrule
\textsc{mean}   &  2.968  &  4.908  &  4.006  &  3.889  &  3.823  &  3.054
      &  4.027  &  3.990  &  1.004  &  0.241  &  1.089  &  0.534  &  1.098 \\
\textsc{re}     & -0.011  & -0.018  &  0.002  & -0.028  &  0.006  &  0.018
      &  0.007  & -0.003  &  0.004  & -0.035  &  0.089  &  0.067  &  0.098 \\
\textsc{std}    &  0.149  &  0.103  &  0.045  &  0.149  &  0.103  &  0.045
      &  0.065  &  0.055  &  0.058  &  0.014  &  0.083  &  0.065  &  0.084 \\
\textsc{inf-cri}&  2.677  &  4.707  &  3.918  &  3.596  &  3.621  &  2.965
      &  3.898  &  3.882  &  0.890  &  0.214  &  0.926  &  0.405  &  0.934 \\
\textsc{sup-cri}&  3.259  &  5.110  &  4.095  &  4.181  &  4.026  &  3.143
      &  4.155  &  4.097  &  1.118  &  0.269  &  1.251  &  0.662  &  1.262 \\
\midrule
\end{tabular}
\begin{tabular}{lrrrr rrr}
   & $\omega_{11}$ & $\omega_{12}$ & $\omega_{13}$ & $\omega_{21}$ & $\omega_{22}$
   & $\omega_{23}$ & \\ \midrule
\textsc{true}
                 &  1.5  &  0.75  &  0.3  &  1.5  &  1.05  &  0.45 & \\ \midrule
\textsc{mean}    &  1.468  &  0.771  &  0.306  &  1.487  &  1.066  &  0.442 & \\
\textsc{re}      & -0.021  &  0.029  &  0.019  & -0.009  &  0.015  & -0.017 & \\
\textsc{std}     &  0.151  &  0.099  &  0.050  &  0.151  &  0.099  &  0.050 & \\
\textsc{inf-cri} &  1.172  &  0.577  &  0.208  &  1.191  &  0.871  &  0.344 &\\
\textsc{sup-cri} &  1.765  &  0.966  &  0.403  &  1.784  &  1.261  &  0.540 & \\
\bottomrule
\end{tabular}
\label{Case1:Sim1MFVB}
\end{table}

\begin{landscape}
\begin{table}
\centering \small \setlength{\tabcolsep}{3pt} \setlength{\extrarowheight}{1.5pt}
\caption{Summary table showing relative errors of the parameters for all simulations
from the frequentist estimation of SUR model, Bayesian estimation of SURME model,
and MFVB estimation of SURME model.}
\begin{tabular}{llrrrr rrrrr rrrrr}
\toprule
&&& $\beta_{11}$     & $\beta_{12}$     & $\beta_{13}$  & $\beta_{21}$  & $\beta_{22}$ & $\beta_{23}$
& $\gamma_1$ & $\gamma_2$ &  $\sigma^{2}_{Z}$ & $\sigma^{2}_{u}$ & $\sigma_{11}$ & $\sigma_{12}$ & $\sigma_{22}$ \\
\midrule
   &   \textsc{sur}       &&  0.381  &  0.124  &  0.062  &  0.288  &  0.223  &  0.118 & -0.198
                          & -0.196  &         &         &  3.171  &  0.046  &  3.144  \\
$(\sigma_{Z}^{2}=1, R_{Z}=0.8)$
   &   \textsc{surme}     && -0.035  & -0.028  & -0.004  & -0.071  & -0.028  &  0.000 &  0.021
                          &  0.026  & -0.026  &  0.005  &  0.025  &  0.008  &  0.015  \\
   &   \textsc{mfvb}      && -0.011  & -0.018  &  0.002  & -0.028  &  0.006  &  0.018 &  0.007
                          & -0.003  &  0.004  & -0.035  &  0.089  &  0.067  &  0.098 \\
\midrule
   &    \textsc{sur}      &&  0.386  &  0.118  &  0.059  &  0.282  &  0.213  &  0.113 & -0.195
                          & -0.190  &         &         &  0.195  &  0.002  &  0.187 \\
$(\sigma_{Z}^{2}=0.0625, R_{Z}=0.8)$
     & \textsc{surme}     && -0.294  & -0.096  & -0.045  & -0.256  & -0.186  & -0.099 &  0.150
                          &  0.168  & -0.123  &  0.433  & -0.079  & -0.008  & -0.090  \\
     & \textsc{mfvb}      && -0.170  & -0.056  & -0.025  & -0.111  & -0.077  & -0.039 &  0.086
                          &  0.070  & -0.049  &  0.174  & -0.026  &  0.011  & -0.027 \\
\midrule
&          \textsc{sur}   &&  0.834  &  0.263  &  0.130  &  0.627  &  0.481  &  0.255 & -0.427
                          & -0.426  &         &         &  6.796  &  0.088  &  6.793 \\
$(\sigma_{Z}^{2}=1, R_{Z}=0.5714)$
   &   \textsc{surme}     && -0.065  & -0.078  & -0.013  & -0.151  & -0.070  & -0.003 &  0.053
                          &  0.058  & -0.081  &  0.010  &  0.023  &  0.005  &  0.020  \\
   &   \textsc{mfvb}      && -0.001  & -0.046  &  0.004  & -0.064  &  0.004  &  0.039 &  0.012
                          & -0.002  &  0.000  & -0.016  &  0.081  &  0.069  &  0.097 \\
\midrule
     & \textsc{sur}       &&  0.844  &  0.256  &  0.128  &  0.625  &  0.468  &  0.251 & -0.425
                          & -0.419  &         &         &  0.424  &  0.000  &  0.413  \\
$(\sigma_{Z}^{2}=0.0625, R_{Z}=0.5714)$
     & \textsc{surme}     && -0.317  & -0.107  & -0.050  & -0.283  & -0.205  & -0.107 &  0.164
                          &  0.184  & -0.154  &  0.154  & -0.070  & -0.023  & -0.087  \\
     & \textsc{mfvb}      && -0.166  & -0.057  & -0.025  & -0.109  & -0.072  & -0.035 &  0.085
                          &  0.066  & -0.055  &  0.042  & -0.019  &  0.003  & -0.016 \\
   \bottomrule
\end{tabular}
\label{Table:SummaryRelErrors}
\end{table}
\end{landscape}

\begin{table}[!t]
\centering \small \setlength{\tabcolsep}{8pt}\setlength{\extrarowheight}{1.5pt}
\setlength\arrayrulewidth{1pt} \caption{Health application - variable
definitions and data summary.}
\begin{tabular}{l lp{8cm}l rrr rrr r }
\toprule
\textsc{variable}   & &   \textsc{description} & \textsc{mean} & \textsc{std}   \\
\midrule
$\ln$(\emph{weight})& &
    Logarithm of weight (in kilograms).           & $4.40$   &  $ 0.24$    \\
\emph{HDL}          & &
    High density lipoprotein (mmol/l, millimoles per litre).
                                                  & $1.32$   &  $ 0.45$    \\
$\ln$(\emph{age})   & &
    Logarithm of age (in years).                  & $4.00$   &  $ 0.15$    \\
\emph{Male}         & &
    Indicator variable for male.                  & $0.57$   &  $ 0.49$    \\
\emph{Smokers}    & &
    Indicator variable for individuals who smoke daily or some days.
                                                  & $0.46$   &  $ 0.50$    \\
\emph{Sedentary}  & &
    Number of hours of sedentary activities.      & $5.28$   &  $ 3.37$    \\
\emph{Sleep disorder}  & &
    Indicator variable for sleep disorder problem.
                                                  & $0.11$   &  $ 0.32$    \\
\emph{LDL20T}  & &
    Low density lipoprotein plus 20 percent of Triglyceride
    (mmol/l, millimoles per litre).
                                                  & $3.94$   &  $ 1.09$    \\
\emph{$\ln$(SBP-50)}   & &
    Transformation on systolic blood pressure.
                                                  & $4.28$   &  $ 0.24$    \\
\bottomrule
\end{tabular}
\label{Table:AppDescStat}
\end{table}

\begin{table}[!t]
\centering \small \setlength{\tabcolsep}{4pt}
\setlength{\extrarowheight}{1.5pt}
\caption{Health application - Bayesian estimation of SUR model. The table
presents the posterior mean (\textsc{mean}), standard deviation (\textsc{std}),
inefficiency factor (\textsc{if}), and Geweke's convergence diagnostics
(\textsc{cd}) of the parameters. Draws=$450,000$, Burnin draws=$50,000$,
thinning=$100$.}

\begin{tabular}{l rrrr r rrrr}
\toprule
& \multicolumn{4}{c}{$\ln(weight)$}   & &  \multicolumn{4}{c}{$HDL$} \\
 \cmidrule{2-5} \cmidrule{7-10}
& \textsc{mean}    &  \textsc{std}     & \textsc{if} & \textsc{cd}
& & \textsc{mean}  &  \textsc{std}     & \textsc{if} & \textsc{cd} \\
\midrule
Intercept & -3.4956  &  0.5894    &  0.9850  &  0.5770 &
          &  1.1074  &  0.3665    &  0.9995  &  0.1262 \\
$\ln(age)$
          & -0.0497  &  0.0443    &  1.0089  & -0.3950 &
          &  0.1035  &  0.0880    &  0.9869  &  0.4550 \\
\emph{Male}
          &  0.0406  &  0.0154    &  0.9914  & -0.8753 &
          & -0.2794  &  0.0264    &  0.9930  & -1.0020 \\
\emph{Smokers}
          & -0.1008  &  0.0132    &  0.9908  &  0.5216 &
          &  0.0424  &  0.0275    &  0.9870  &  1.0594 \\
\emph{Sedentary}
          &  0.0085  &  0.0019    &  0.9821  &  -0.2897 &
          & -0.0076  &  0.0039    &  1.0069  &  -0.4709 \\
\emph{Sleep disorder}
          &  0.1061  &  0.0202    &  1.0197  &  -0.1130 &
          & -0.0272  &  0.0412    &  0.9903  &   0.2408 \\
\emph{LDL20T}
          &  0.0223  &  0.0058    &  1.0098  &  -0.4155 &
          & -0.1137  &  0.0119    &  1.0024  &  -0.1327 \\
$\ln(height)$
          &  1.4731  &  0.1134    &  0.9949  &  -0.6200 &
          & & & & \\
$\ln(SBP-50)$
          &  0.0967  &  0.0277    &  0.9963  &   0.8784 &
          &  0.1012  &  0.0560    &  0.9873  &  -0.7712 \\
$\sigma_{11}$
          &  0.0399  &  0.0018    &  0.9900  &  -1.0924 &
          & & & & \\
$\sigma_{12}$
          & -0.0250  &  0.0027    &  1.0219  &   0.4914 &
          & & & &  \\
$\sigma_{22}$
          &  0.1693  &  0.0075    &  1       &  -0.5193 &
          & & & & \\ \bottomrule
\end{tabular}
\label{Table:AppBSUR}
\end{table}

\begin{table}[!t]
\centering \small \setlength{\tabcolsep}{4pt}
\setlength{\extrarowheight}{1.5pt}
\caption{Health application - Bayesian
estimation of SURME model. The upper panel presents results for the main
equations and the lower panel presents results for the exposure equations.
\textsc{mean} is posterior mean, \textsc{std} is posterior
standard deviation, \textsc{if} is inefficiency factor, and \textsc{cd}
is Geweke's convergence diagnostics.
Draws=$450,000$, Burnin draws=$50,000$, thinning=$100$.}
\begin{tabular}{l rrrr r rrrr}
\toprule
& \multicolumn{4}{c}{$log(weight)$}  & & \multicolumn{4}{c}{$HDL$} \\
\cmidrule{2-5} \cmidrule{7-10}
& \textsc{mean}    &  \textsc{std}   & \textsc{if} & \textsc{cd}
& & \textsc{mean}  &  \textsc{std}   & \textsc{if} & \textsc{cd} \\
\midrule
Intercept  &  -5.2859  &  0.8284    &  1.0132  &  0.6022  &
           &   1.1825  &  0.4332    &  1.7579  &  0.7772 \\
$\ln(age)$
           &  -0.0383  &  0.0521    &  1.1769  &  0.7601  &
           &   0.0329  &  0.0997    &  1.4330  &  1.7639 \\
\emph{Male}
           &   0.0156  &  0.0187    &  1.0181  & -0.0596  &
           &  -0.2807  &  0.0276    &  1.0156  &  0.7971 \\
\emph{Smokers}
           &  -0.1009  &  0.0146    &  0.9998  &  0.8830  &
           &   0.0375  &  0.0280    &  1.0110  &  1.7027 \\
\emph{Sedentary}
           &   0.0080  &  0.0021    &  1.0279  &  0.8739 &
           &  -0.0075  &  0.0040    &  1.0071  & -0.1389 \\
\emph{Sleep disorder}
           &   0.1066  &  0.0222    &  0.9863  &  0.8386 &
           &  -0.0294  &  0.0425    &  0.9941  & -0.0086 \\
\emph{LDL20T}
           &   0.0224  &  0.0066    &  1.0159  &  1.4224 &
           &  -0.1155  &  0.0126    &  1.0008  & -0.1091 \\
$\ln(height)$
           &   1.7790  &  0.1561    &  1.0112  & -0.4465 &
           & & & & \\
$\ln(SBP-50)$
           &   0.1416  &  0.0443    &  3.3261  & -0.6761 &
           &   0.1520  &  0.0921    &  3.8691  & -1.1416 \\
$\sigma_{11}$
           &   0.0485  &  0.0022    &  0.9933  &  0.1713 &
           & & & &  \\
$\sigma_{12}$
           &  -0.0252  &  0.0030    &  0.9829  & -0.7013 &
           & & & &  \\
$\sigma_{22}$
           &   0.1778  &  0.0080    &  1.0163  & -0.3718 &
           & & & &  \\
$\sigma^{2}_{Z}$
           &   0.0440  &  0.0019    &  1.0025  &  0.9275 &
           & & & &  \\
$\sigma^{2}_{u}$
           &   0.0296  &  0.0016    &  1.0275  & -1.4186 &
           & & & & \\
\midrule
& \multicolumn{4}{c}{$\ln(SBP-50)$} & & \multicolumn{4}{c}{$\ln(SBP-50)$} \\
\cmidrule{2-5} \cmidrule{7-10}
& \textsc{mean}    &  \textsc{std}  & \textsc{if} & \textsc{cd}
& & \textsc{mean}  &  \textsc{std}  & \textsc{if} & \textsc{cd} \\
\midrule
Intercept
    &  1.1847  &  0.7160    &  1.0137  &  -0.0465  &
    &  2.4526  &  0.2377    &  1.0053  &   0.8891  \\
$\ln(age)$
    &  0.4156  &  0.0577    &  1.0250  &   0.8542  &
    &  0.4302  &  0.0580    &  1.0051  &  -0.8192  \\
\emph{Male}
    &  0.0118  &  0.0201    &  0.9935  &  -0.0731  &
    &  0.0318  &  0.0172    &  1.0165  &  -0.4137 \\
\emph{Smokers}
    &  0.0341  &  0.0180    &  1.0020  &   0.6742  &
    &  0.0363  &  0.0179    &  1.0280  &  -1.3321  \\
\emph{Sedentary}
    & -0.0018  &  0.0026    &  0.9709  &  0.1812 &
    & -0.0013  &  0.0025    &  0.9886  &  1.3253 \\
\emph{Sleep disorder}
    &  0.0087  &  0.0277    &  0.9841  & -0.2518 &
    &  0.0094  &  0.0274    &  1.0220  &  0.8595 \\
\emph{LDL20T}
    &  0.0185  &  0.0079    &  1.0111  & -0.1604 &
    &  0.0190  &  0.0080    &  1.0092  & -0.8225 \\
$\ln(height)$
    &  0.2620  &  0.1405    &  1.0129  &-0.2297  &
    & & & &  \\
\bottomrule
\end{tabular}
\label{Table:AppBSURME}
\end{table}

\begin{table}[!t]
\centering \small \setlength{\tabcolsep}{6pt}
\setlength{\extrarowheight}{1.5pt}
\caption{Health application - MFVB estimation of SURME model. Cycles=$1565$,
Maximum Elbo =$-9076.16$. The upper panel presents results for the main
equations and the lower panel presents results for the exposure equations.
\textsc{mean} is posterior mean and \textsc{std} is posterior
standard deviation of the parameters.}
\begin{tabular}{l rr r rr}
\toprule
& \multicolumn{2}{c}{$\ln(weight)$} & & \multicolumn{2}{c}{$HDL$} \\
\cmidrule{2-3} \cmidrule{5-6}
& \textsc{mean}    &  \textsc{std}   & & \textsc{mean}  &  \textsc{std} \\
\midrule
Intercept              &  -5.3245  &  0.8069  &&  1.0892  &  0.3726 \\
$\ln(age)$             &  -0.0461  &  0.0477  &&  0.0199  &  0.0908 \\
\emph{Male}            &   0.0148  &  0.0185  && -0.2816  &  0.0272 \\
\emph{Smokers}         &  -0.1018  &  0.0144  &&  0.0364  &  0.0276 \\
\emph{Sedentary}       &   0.0080  &  0.0021  && -0.0075  &  0.0040 \\
\emph{Sleep disorder}  &   0.1066  &  0.0222  && -0.0293  &  0.0425 \\
\emph{LDL20T}          &   0.0222  &  0.0064  && -0.1160  &  0.0123 \\
$\ln(height)$          &   1.7781  &  0.1537  &&          &  \\
$\ln(SBP-50)$          &   0.1594  &  0.0422  &&  0.1867  &  0.0808 \\
$\sigma_{11}$          &   0.0484  &  0.0022  &&          &  \\
$\sigma_{12}$          &  -0.0252  &  0.0030  &&          &  \\
$\sigma_{22}$          &   0.1778  &  0.0079  &&          &  \\
$\sigma^{2}_{Z}$       &   0.0443  &  0.0014  &&          &  \\
$\sigma^{2}_{u}$       &   0.0293  &  0.0009  &&          &  \\
\midrule
& \multicolumn{2}{c}{$\ln(SBP-50)$} & & \multicolumn{2}{c}{$\ln(SBP-50)$} \\
\cmidrule{2-3} \cmidrule{5-6}
& \textsc{mean}    &  \textsc{std}   & & \textsc{mean}  &  \textsc{std} \\
\midrule
Intercept              &  1.1679  &  0.6340  &&  2.4525  &  0.1843 \\
$\ln(age)$             &  0.4153  &  0.0451  &&  0.4305  &  0.0449 \\
\emph{Male}            &  0.0118  &  0.0165  &&  0.0318  &  0.0136 \\
\emph{Smokers}         &  0.0340  &  0.0138  &&  0.0362  &  0.0138 \\
\emph{Sedentary}       & -0.0018  &  0.0020  && -0.0012  &  0.0020 \\
\emph{Sleep disorder}  &  0.0090  &  0.0213  &&  0.0095  &  0.0213 \\
\emph{LDL20T}          &  0.0186  &  0.0061  &&  0.0187  &  0.0061 \\
$\ln(height)$          &  0.2655  &  0.1228  &&          & \\
\bottomrule
\end{tabular}
\label{Table:AppMFVB}
\end{table}

\newpage

\end{document}